\documentclass[longauth]{aa} 

\usepackage{graphicx}
\usepackage{txfonts}
\usepackage{natbib}
\bibpunct{(}{)}{;}{a}{}{,} 

\begin{document} 

   \title{First Light for GRAVITY: Phase Referencing Optical Interferometry for the Very Large Telescope Interferometer}
   \titlerunning{First Light for GRAVITY}

   \author{
GRAVITY Collaboration\thanks{GRAVITY is developed in a collaboration by the Max Planck Institute for extraterrestrial Physics, LESIA of Paris Observatory / CNRS / UPMC / Univ. Paris Diderot and IPAG of Universit\'e Grenoble Alpes / CNRS, the Max Planck Institute for Astronomy, the University of Cologne, the Centro Multidisciplinar de Astrofísica Lisbon and Porto, and the European Southern Observatory.}:
R.~Abuter \inst{8} \and 
M.~Accardo \inst{8} \and 
A.~Amorim \inst{6} \and 
N.~Anugu \inst{7} \and 
G.~Ávila \inst{8} \and 
N.~Azouaoui \inst{2} \and 
M.~Benisty \inst{5} \and
J.P.~Berger \inst{5} \and
N.~Blind \inst{10} \and 
H.~Bonnet \inst{8}
% ...
}

   \authorrunning{GRAVITY Collaboration} 
   
   \institute
   {    
                 Max Planck Institute for extraterrestrial Physics, Giessenbachstr., 85748 Garching, Germany
         \and
                 LESIA, Observatoire de Paris, PSL Research University, CNRS, Sorbonne Universit\'es, UPMC Univ. Paris 06, Univ. Paris Diderot, Sorbonne Paris Cit\'e, France
         \and
                 Max-Planck-Institut für Astronomie, Königstuhl 17, 69117 Heidelberg, Germany
         \and
                 1. Physikalisches Institut, Universität zu Köln, Zülpicher Str. 77, 50937 Köln, Germany
         \and
                 Univ. Grenoble Alpes, CNRS, IPAG, F-38000 Grenoble, France
         \and
                 CENTRA and Universidade de Lisboa - Faculdade de Ciências, Campo Grande, 1749-016 Lisboa, Portugal
         \and
                 CENTRA and Universidade do Porto - Faculdade de Engenharia,  4200-465 Porto, Portugal
         \and
                 European Southern Observatory, Karl-Schwarzschild-Str. 2, 85748 Garching, Germany
         \and
                 European Southern Observatory, Casilla 19001, Santiago 19, Chile
         \and
                 Observatoire de Genève, Université de Genève, 51 ch. des Maillettes, 1290 Versoix, Switzerland
         \and  
                 Onera - The French Aerospace Lab, BP 72, 92 322 Châtillon, France
 	    \and 
	         Unidad Mixta Internacional Franco-Chilena de Astronomía (CNRS UMI 3386), Departamento de Astronomía, Universidad de Chile, Camino El Observatorio 1515, Las Condes, Santiago, Chile
         \and
                 Max-Planck-Institute for Radio Astronomy, Auf dem Hügel 69, 53121 Bonn, Germany
         \and  
                 Sterrewacht Leiden, Leiden University, Postbus 9513, 2300 RA Leiden, The Netherlands
         \and
		        Department of Physics, Le Conte Hall, University of California, Berkeley, CA 94720, USA
   }

   \date{Accepted for publication in A\&A, submitted 21 March, 2017, revised 26 April, 2017}
 
  \abstract
   {
GRAVITY is a new instrument to coherently combine the light of the European Southern Observatory Very Large Telescope Interferometer to form a telescope with an equivalent $130\,\text{m}$ diameter angular resolution and a collecting area of $200\,\text{m}^2$. The instrument comprises fiber fed integrated optics beam combination, high resolution spectroscopy, built-in beam analysis and control, near-infrared wavefront sensing, phase-tracking, dual beam operation and laser metrology. GRAVITY opens up to optical/infrared interferometry the techniques of phase referenced imaging and narrow angle astrometry, in many aspects following the concepts of radio interferometry. This article gives an overview of GRAVITY and reports on the performance and the first astronomical observations during commissioning in 2015/16. We demonstrate phase-tracking on stars as faint as $m_K \approx 10$\,mag, phase-referenced interferometry of objects fainter than $m_K \approx 15$\,mag with a limiting magnitude of $m_K \approx 17$\,mag, minute long coherent integrations, a visibility accuracy of better than $0.25\,\%$, and spectro-differential phase and closure phase accuracy better than $0.5^\circ$, corresponding to a differential astrometric precision of better than 10\,microarcseconds ($\mu\text{as}$). The dual-beam astrometry, measuring the phase difference of two objects with laser metrology, is still under commissioning. First observations show residuals as low as $50\,\mu\text{as}$ when following objects over several months. We illustrate the instrument performance with the observations of archetypical objects for the different instrument modes. Examples include the \,Galactic Center supermassive black hole and its fast orbiting star S2 for phase referenced dual beam observations and infrared wavefront sensing, the High Mass X-Ray Binary BP\,Cru and the Active Galactic Nucleus of PDS\,456 for few $\mu\text{as}$ spectro-differential astrometry, the T\,Tauri star S\,CrA for a spectro-differential visibility analysis, $\xi$\,Tel and 24\,Cap for high accuracy visibility observations, and $\eta$ Car for interferometric imaging with GRAVITY.  
   }
  

   \maketitle


\section{Introduction}

\subsection{From double slit to phase referenced interferometry}

About 150 years ago, \cite{Fizeau1868} introduced the concept of stellar interferometry as a double slit experiment. This double slit technique allowed \cite{Stephan1874} to derive strong upper limits for the diameter of stars and was then brought to full fruition by \cite{1921ApJ....53..249M} with the measurement of the diameter of Betelgeuse. More than 50 years later, \cite{1975ApJ...196L..71L} was able to demonstrate the interference between two telescopes as the basis for modern optical interferometry. At this time radio-interferometry was already well advanced with first imaging synthesis arrays and with Very Long Baseline Interferometry (VLBI) for highest angular resolution and astrometry (see, e.g., \citealt{2017isra.book.....T} and references therein). The discoveries from radio interferometry  -- e.g., the imagery of Cygnus A \citep{1974MNRAS.166..305H} and the observations of apparent superluminal motion in 3C273 \citep{1981Natur.290..365P} -- also set the directions for optical/infrared interferometry. But it needed significant technical advances in technology  -- to name a few: detectors, optics, electronics, computers and lasers -- to arrive at modern optical/infrared interferometers, e.g., the Mark III stellar interferometer \citep{1988A&A...193..357S}. 

The currently largest optical/infrared interferometers are CHARA \citep{2005ApJ...628..453T} with six 1\,m diameter  telescopes, and the  European Southern Observatory (ESO) Very Large Telescope Interferometer (VLTI,  \citealt{2012SPIE.8445E..0DH}), combining either up to four 8\,m diameter Unit Telescopes (UTs) or up to four movable 1.8\,m diameter Auxiliary Telescopes (ATs). In comparison with interferometers at radio wavelengths, optical/near-infrared interferometry is penalized by three fundamental limitations: (1) the practical and fundamental limitations for a broad band heterodyne detection -- pioneered and applied for wavelengths down to 10\,$\mu$m with the Infrared Spatial Interferometer (ISI) and its prototype \citep{1974PhRvL..33.1617J,2000ApJ...537..998H} --, and the atmospheric turbulence leading to (2) only partial coherence of the beams of each telescope, and to (3) short coherence times for the interference between the telescopes. 

The effect of the variable wavefront coherence of each telescope is notoriously difficult to cope with at shorter wavelengths. This problem was overcome in the 1990's with the introduction of single-mode fibers at the FLUOR interferometer \citep{1998SPIE.3350..856C}, which spatially filter the wavefronts corrugated by the atmospheric turbulence, and allow for visibility accuracies as good as a few 0.1\,\% \citep{2004A&A...426..279P}. In case of large telescope, adaptive optics is necessary to maximize the coupling to the single mode fibers. This has first been done at the VLTI with the Multi-Application Curvature Adaptive Optics (MACAO, \citealt{2003SPIE.4839..174A}).

The second problem of atmospheric turbulence, the short coherence time between two telescopes, limits the detector integration time to typically less than 100\,ms in the astronomical K-band ($1.95-2.45\,\mu$m), and because of detector read noise, to objects brighter than $m_K \approx 10$\,mag for broad band observation even for 10\,m class telescopes \citep{2012A&A...541L...9W,2011A&A...527A.121K}, and significantly brighter objects for high spectral resolution. The technology to stabilize the phase between two telescopes to a fraction of a wavelength is called phase-tracking or fringe-tracking\footnote{The term fringe-tracking is also used in a wider sense for stabilizing the optical path difference (OPD) between the telescopes to within a coherence length, which depending on the spectral resolution is several to many wavelengths.}. Phase-tracking was first demonstrated by \cite{1980ApOpt..19.1519S}, used on the Mark III interferometer \citep{1988A&A...193..357S} and the Palomar Testbed Interferometer (PTI, \citealt{1999ApJ...510..505C}), and later, e.g., at the Keck Interferometer (KI, \citealt{2003SPIE.4838...79C}) and with the FINITO fringe tracker \citep{2003MmSAI..74..472G} at the VLTI. Because these closed loop systems require significantly shorter detector integration times of typically a few ms, the broad-band sensitivity itself does not improve, but fringe-tracking boosts the sensitivity for parallel high-spectral resolution observations with long, coherent integrations. Fringe-tracking also opens up the possibility of observing objects close to a reference star with long coherent exposures, greatly increasing also the broad band limiting magnitude, and at the same time providing the phase reference for precise narrow angle astrometry \citep{1992A&A...262..353S}, visibility measurements \citep{1994A&A...286.1019Q}, and radio-VLBI-like imaging \citep{1989ASIC..283..261A}. Such off-axis fringe-tracking was implemented in PTI \citep{1999ApJ...510..505C}, the Dual Field Phase Referencing instrument \citep{2014ApJ...783..104W} for the KI, and PRIMA \citep{2008NewAR..52..199D} for the VLTI. These instruments demonstrated the potential of this technique with $10\,\mu$as-astrometry \citep{2004ApJ...601.1129L} and interferometry of objects as faint as $m_K \approx 12.5$\,mag \citep{2014ApJ...783..104W}, but to date have not yet been exploited scientifically. 

\subsection{From astrophysical questions to GRAVITY}

Inspired by the potential of phase-referenced interferometry to zoom in onto the black hole in the Galactic Center and to probe its physics down to the event horizon \citep{2008poii.conf..313P}, we proposed in 2005 a new instrument named GRAVITY as one of the second generation VLTI instruments \citep{2008poii.conf..431E}. At its target accuracy and sensitivity, GRAVITY will also map with spectro-differential astrometry the broad line regions of Active Galactic Nuclei (AGN), image circumstellar discs in young stellar objects and see their jets evolve in real-time, and detect and characterize exo-planets especially around low mass stars and binaries -- in short we will ``Observe the Universe in motion'' \citep{2011Msngr.143...16E}. 

The instrument derives its design and optimization from focusing on the science themes mentioned above: K-band operation -- both wavefront sensor and beam combiner -- for optimum resolution and sensitivity in highly dust-extincted regions and access to the most important near-infrared line diagnostics including Br$\gamma$, \ion{He}{i,ii}, H$_2$, and the CO band heads; fringe-tracking with a limiting magnitude fainter than $m_K \approx 10$\,mag for the Galactic Center phase reference stars and giving access to the brightest AGN and low-mass T\,Tauri stars; a broad band limiting magnitude fainter than $m_K \approx 16$\,mag -- for off-axis fringe-tracking on a bright reference star -- to trace flares around the Galactic Center black hole and the fast motions of stars at distances smaller than 100\,mas; a medium $R = \lambda/\Delta\lambda \approx 500$ and high $R \approx 4500$ spectral resolving power to probe extragalactic and circumstellar velocities, respectively; and narrow angle astrometry well below 100\,$\mu$as with a goal of 10\,$\mu$as to probe general relativistic effects around the Galactic Center black hole and to potentially detect planetary mass companions around nearby stars. 

Following a one-year phase-A study, the instrument was selected at the end of 2007. We presented the preliminary and final designs in 2009/10 and 2011/12, respectively. The beam-combiner instrument was shipped to Paranal in mid 2015 with the four infrared wavefront sensors following between February and July 2016. 
  
\begin{figure*}
\vspace{0 cm}
\centering
\includegraphics[width=\hsize]{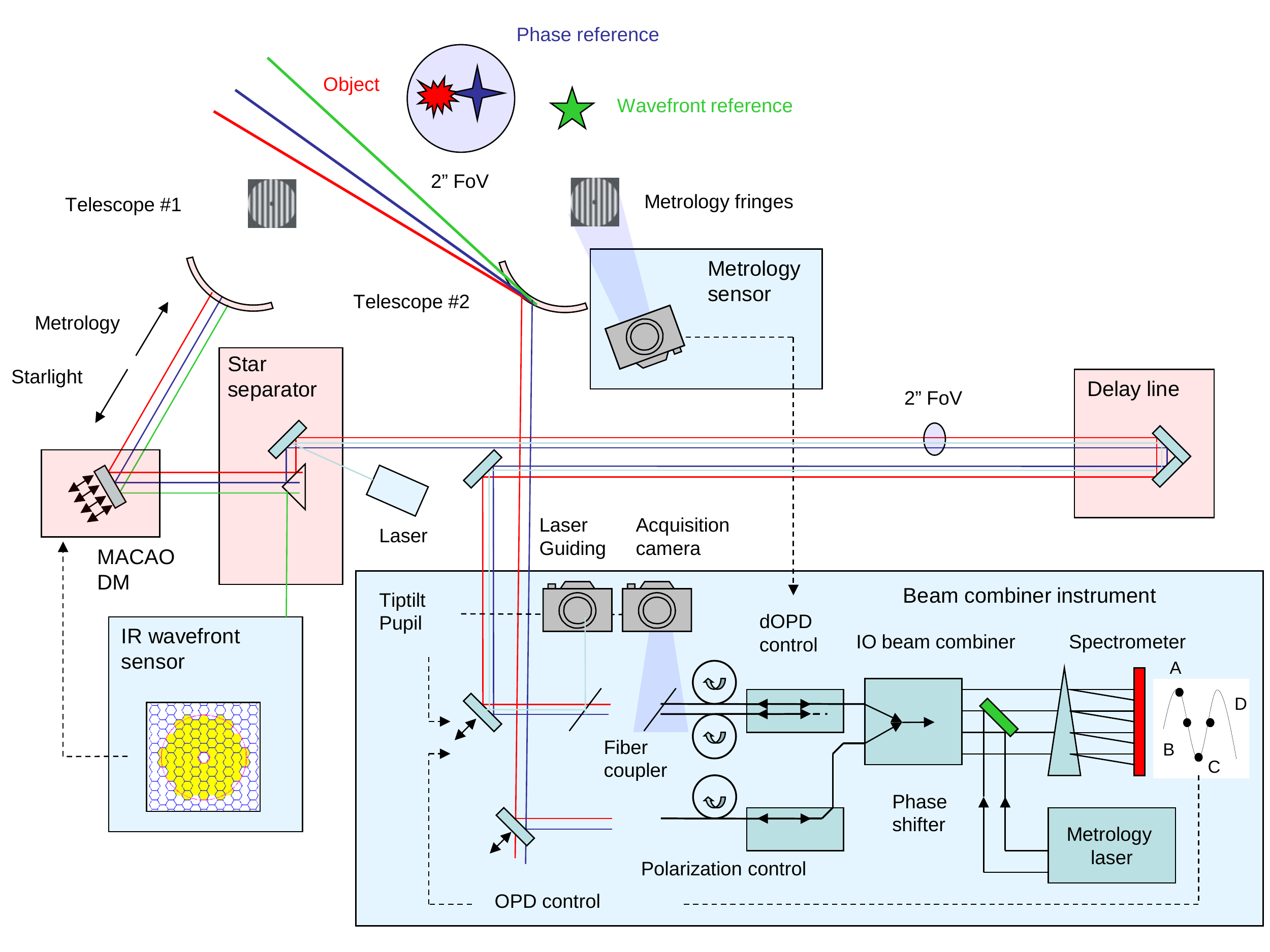}
\vspace{0 cm}
\caption{Overview and working principle of GRAVITY: the instrument coherently combines the light of the four 8m UTs of the VLTI or the four 1.8m ATs. It provides infrared wavefront sensing to control the telescope adaptive optics, two interferometric beam combiners - one for fringe-tracking and one for the science object -, an acquisition camera and various laser guiding systems for beam stabilization, and a dedicated laser metrology to trace the optical path length differences for narrow angle astrometry. The overview illustrates the light path and location of the various subsystems. For clarity, we show only two telescopes, one beam combiner and one wavefront sensor. The wavefront sensor star (green) can be outside the VLTI field of view. Depending on the brightness of the science object (red), the fringe tracker is either fed by a beam splitter, or - as illustrated in this figure - by a bright off-axis phase reference star (blue). The GRAVITY subsystems (light blue boxes) are embedded and take advantage of the already existing VLTI infrastructure (light red boxes). }
\label{fig:Overview}
\end{figure*}

\subsection{First light for GRAVITY}

One year after the start of the VLTI upgrade for the $2^\text{nd}$ generation instruments \citep{2016SPIE.9907E..1ZG}, we began the commissioning of the GRAVITY beam combiner instrument with the four ATs in November 2015\footnote{\url{http://www.eso.org/public/news/eso1601/}}. The commissioning with the four UTs started in May 2016. The first infrared wavefront sensor saw first starlight in April 2016. The full GRAVITY instrument with all infrared wavefront sensors had its first observations with the four UTs in September 2016. Science verification -- following proposals from the community and with open data access -- with the ATs was carried out in June and September 2016, with first results published in, e.g.,  \cite{2016arXiv160803525L} and \cite{2017ApJ...835L...5K}. Science operation with the ATs started in October 2016, followed by the UTs in April 2017.

This paper provides a comprehensive description of the instrument (Section \ref{sec:TheGRAVTITYInstrument}) and presents a set of early observations that illustrate its power (Section \ref{sec:FirstGRAVITYobservations}). The detailed description of the instrument subsystems and software, and of the analysis and interpretation of the observations will be given in several forthcoming papers.


\section{The GRAVITY instrument}
\label{sec:TheGRAVTITYInstrument}

\subsection{Overview and working principle}

The goal of the GRAVITY design is to provide a largely self-contained instrument for phase-referenced imaging of faint targets and precise narrow angle astrometry. Fig.~\ref{fig:Overview} illustrates the GRAVITY concept. For clarity, only two of four telescopes, i.e. one out of six baselines, are shown. 

The working principle of GRAVITY is as follows: a bright wavefront reference star (e.g., in the Galactic Center this is GC\,IRS\,7, a $m_K = 6.5$\,mag star at 5.5\arcsec\ separation from the supermassive black hole) outside the 2\arcsec\ field-of-view of the VLTI is picked with the PRIMA star separator \citep{2004SPIE.5491.1528D} and imaged onto the GRAVITY Coud\'e Infrared Adaptive Optics (CIAO) wavefront sensors. The wavefront correction is applied using the MACAO deformable mirrors of the UTs. The 2\arcsec\ field-of-view of the VLTI contains both the science target (Sgr\,A*) and the phase reference star (GC\,IRS\,16C, 1.23\arcsec\ separation, $m_K = 9.7$\,mag).  Both objects are re-imaged via the main delay lines \citep{2000SPIE.4006...25D} to the GRAVITY beam combiner instrument. Laser guiding beams are launched at the star separator and telescope spider arms to trace the tip-tilt and pupil motion, respectively, within the VLTI beam relay. The GRAVITY beam combiner instrument has internal sensors and actuators to analyze these beams and to apply the corresponding corrections. Longer-term image drifts of the object are compensated with the help of the internal acquisition camera (working at H-band, $1.45-1.85\,\mu$m). This camera also analyses the signal from the pupil-guiding laser beams launched at the telescope spider arms. The fiber coupler de-rotates the field, splits the light of the two stars and injects it into single-mode fibers. A rotating half-wave plate is used to control the linear polarization of the light. A fiber control unit including rotators and stretchers aligns the polarization for maximum contrast, and compensates the differential OPD between the phase reference star and science object caused by their angular separation on sky. The beam combiner itself is implemented as an integrated optics chip with instantaneous fringe sampling. The bright reference star feeds the fringe tracker, which measures the phase and group delay from six spectral channels across the K-band. The OPD correction is applied to an internal piezo-driven mirror stabilizing the fringes of both the reference star and the faint science object. The science spectrometer is optimized for longer, background-limited integration times of faint objects, and offers a variety of operation modes, including broad band (10 spectral pixel) observations and $R \approx 500$ and $R \approx 4500$ resolution spectroscopy. Both the fringe tracker and the science spectrometer can be used with a Wollaston prism to split and simultaneously measure two linear polarization states. The differential OPD (dOPD) between the science and reference beams is measured with a laser metrology system. The laser light is back-propagated from the GRAVITY beam combiners covering the full beam up to above the telescope primary mirror. The metrology is implemented via phase-shifting three-beam interferometry and measured by photodiodes mounted on the telescope spider arms. A dedicated calibration unit simulates the light from two stars and four telescopes, and provides all functions to test and calibrate the beam combiner instrument. 

GRAVITY provides simultaneously for each spectral channel the visibility of the reference and science object, and the differential phase between reference and science object. The GRAVITY data can be used for interferometric imaging exploring visibilities and closure phases obtained simultaneously for six baselines, and for astrometry using the differential phases and group delays. The spatial frequency coverage can be further extended taking advantage of the earth rotation, and in the case of the ATs, with the relocation of the telescopes. 

\subsection{Beam combiner instrument}

\begin{figure}
\vspace{0 cm}
\centering
\includegraphics[width=\hsize]{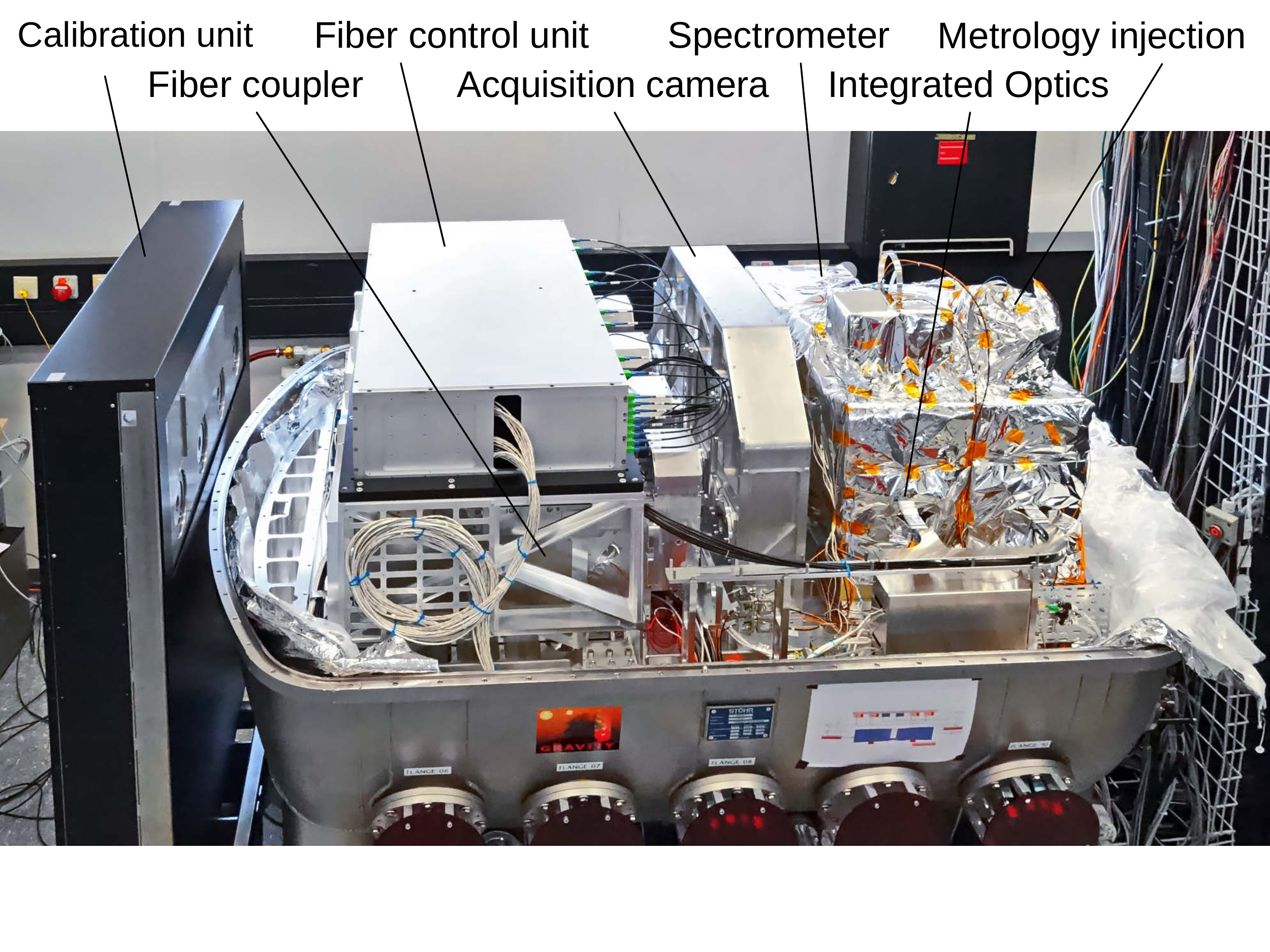}
\vspace{-0.5 cm}
\caption{Beam combiner instrument: the photograph shows the instrument with the vacuum vessel removed to expose the subsystems. The light from the four telescopes enters from the left.  The fiber couplers are located in the left part of the instrument below the fiber control unit. The acquisition camera and the receivers for the tip/tilt laser stabilization are seen in the middle of the instrument. The black cables are single-mode fibers connecting the fiber control unit with the fiber couplers and the integrated optics beam combiner. These integrated optics beam combiners are mounted to the two spectrometers - wrapped in shiny super isolation - seen on the right. The laser metrology is injected through the little ``huts'' - also wrapped in super isolation - on top of the spectrometers. The warm calibration unit is the black box on the very left. }
\label{fig:BeamCombinerInstrument}
\end{figure}

The beam combiner instrument is installed in the VLTI interferometric laboratory located at the center of the Paranal observatory. Most subsystems are hosted in a cryostat for optimum stability, cleanliness, and thermal background suppression (see Fig.~\ref{fig:BeamCombinerInstrument}). 

\subsubsection{Cryostat}

The cryostat \citep{2012SPIE.8445E..2VH} provides the required temperatures for the various subunits of the beam combiner instrument. The temperatures range from about 80 K for the detectors and spectrometers, 200 K for the integrated optics, 240 K for the optical bench and fiber couplers, and up to 290 K for the metrology injection units. The bath-cryostat is cooled with liquid nitrogen, and makes use of the gaseous exhaust to cool the intermediate 240 K temperature subsystems. All temperature levels are actively stabilized with electric heaters. The cold bench is supported separately from the vacuum vessel and liquid nitrogen reservoir to minimize vibrations within the instrument.

\subsubsection{Fiber coupler}
\label{FiberCoupler}

\begin{figure}
\vspace{0 cm}
\centering
\includegraphics[width=\hsize]{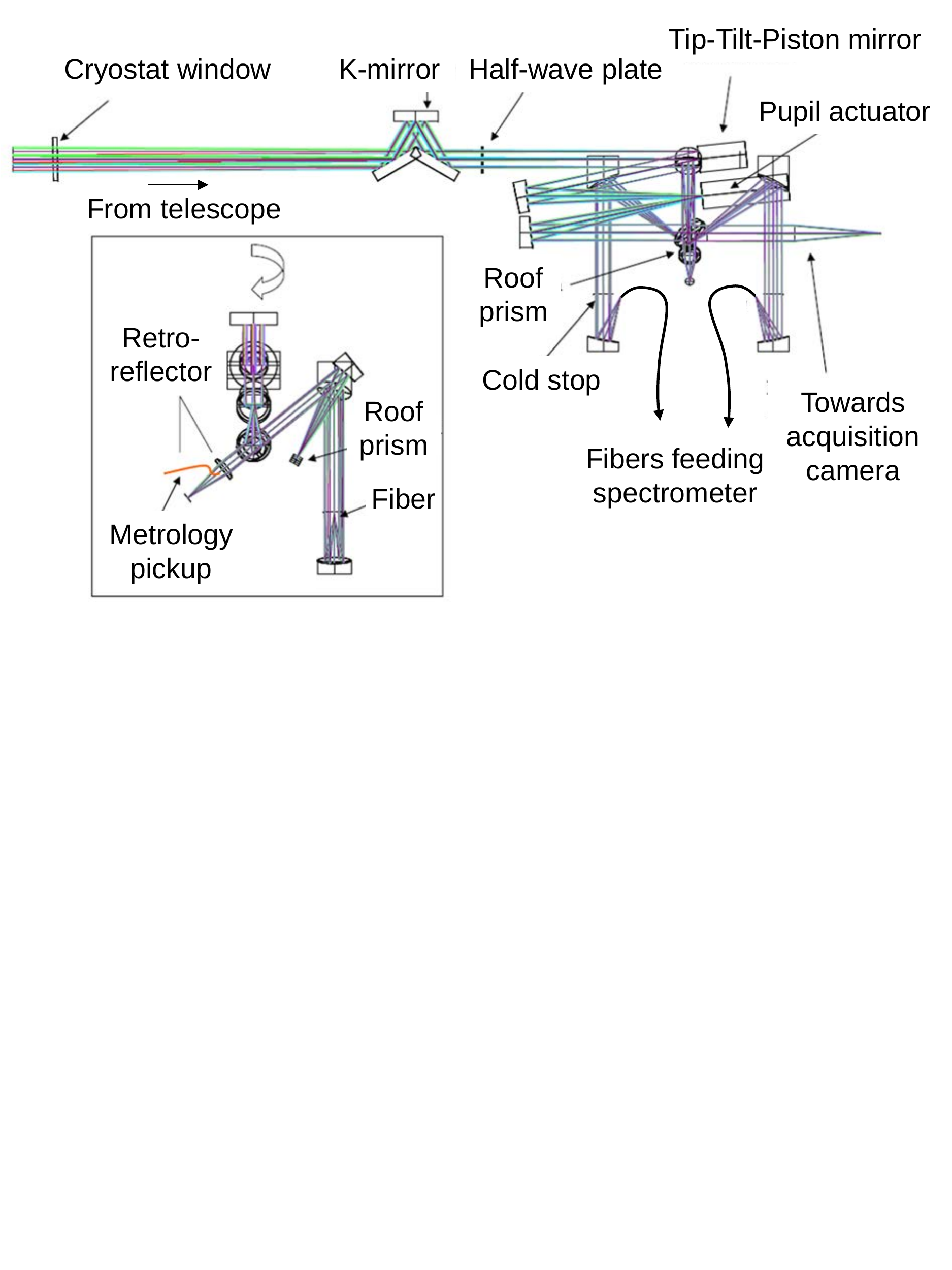}
\vspace{-6 cm}
\caption{Fiber coupler: the schematic shows the side and front views of the optical design for one of the four fiber couplers. The fiber couplers provide all optics and actuators to rotate and stabilize the beam, to split the light of the fringe tracker and science objects,  and to optimally couple the light into the single-mode fibers. In addition, the fiber coupler contains several supplementary functions, including a metrology pickup to servo the fiber differential delay lines, a retro-reflector to image the fiber input on the acquisition  camera, and motorized half-wave plates for co-aligning the polarization.}
\label{fig:FiberCoupler}
\end{figure}

The main purpose of the fiber coupler \citep{2014SPIE.9146E..23P} is to feed the light from the reference and science objects into the fibers. Fig.~\ref{fig:FiberCoupler} shows the optical design for one of the four units. Every fiber coupler provides a number of functions. First, a motorized K-mirror corrects the field-rotation induced by the VLTI optical train. After that, a motorized half-wave plate allows independently rotating the linear polarization. Two piezo driven mirrors provide tip-tilt-piston and lateral pupil control, respectively. They are part of an off-axis parabolic mirror relay optics, which focuses the starlight onto a roof-prism. One part of the roof-prism is fully reflective to completely separate the phase reference and the science star light, another part of the roof is a beam-splitter to send half of the light to the fringe tracker and science spectrometer, respectively. Two separate relay optics then couple the phase reference and science starlight into their respective fibers, which are mounted on piezo-driven three-axis stages to pick the objects and adjust the focus. The acquisition and guiding camera is fed via a dichroic beam splitter. To ease the alignment, a retro-reflector behind the dichroic beam splitter allows imaging the fiber entrance onto the acquisition camera. Behind the dichroic beam splitter, there is also a multi-mode fiber to pick up part of the laser metrology light to servo the fiber differential delay lines. 

\subsubsection{Single-mode fibers and fiber control unit}
\label{SingleModeFibersAndFiberControl}

GRAVITY uses fluoride-glass single-mode fibers to transport the light from the fiber couplers to the integrated optics, where the beams from the four telescopes interfer. The fibers also spatially filter the wavefronts corrugated by the atmospheric turbulence. As such the phase fluctuations are traded against photometric fluctuations, which are measured by the beam combiner to calibrate the coherence losses. GRAVITY uses weakly birefringent fluoride glass fibers (beat lengths between 313\,m and 1074\,m) such that  for the fiber length of 20.5\,m to 22\,m per beam an intrinsic maximum contrast higher than 98.2\,\% can be achieved without splitting polarizations to ensure maximum sensitivity on faint objects. The fiber lengths have been matched to simultaneously minimize optical path differences and differential dispersion between the four beams of the fringe tracker and science beam combiner, respectively. Respective maximum values for each of the six baselines are 0.9\,mm and 5.7\,$\mu$rad\,cm$^2$, which corresponds to a contrast loss of less than 5\,\% for the low spectral resolution mode (see Section \ref{sec:Spectrometer}) of GRAVITY, and negligible contrast losses at medium and high spectral resolution. 

\begin{figure}
\vspace{0 cm}
\centering
\includegraphics[width=\hsize]{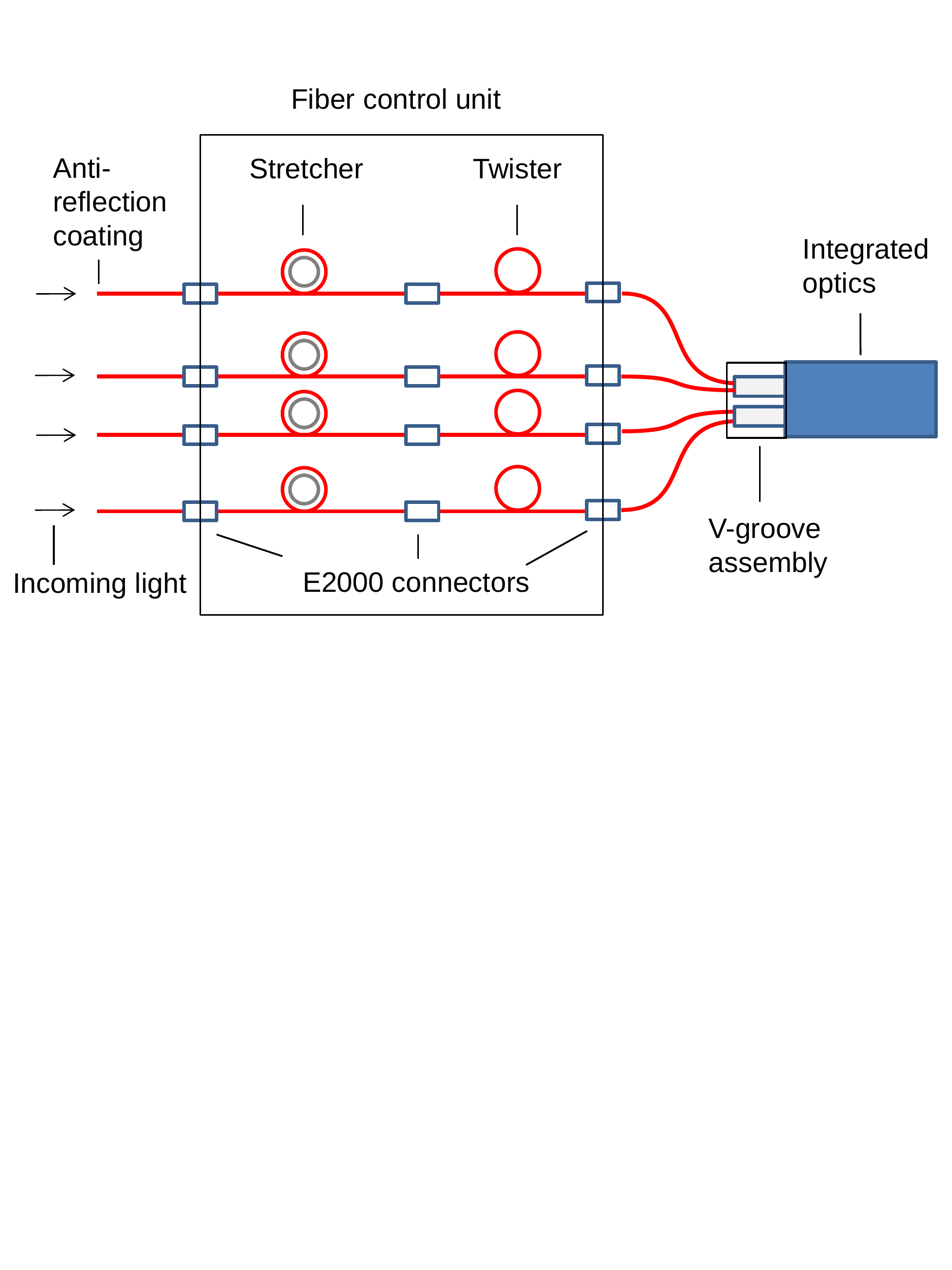}
\vspace{-6 cm}
\caption{Single-mode fibers and fiber control unit: the schematic shows the various fiber subsections and functions. The entrance of the fibers are anti-reflection coated. The optical path length is adjusted by piezo driven fiber coils. The direction of linear polarization is aligned by motorized fiber twisters. The fibers of these subsections are connected through optimized E2000 connectors. The integrated optics beam combiner - operated at 200K - is connected through a dedicated V-groove assembly.}
\label{fig:Fibers}
\end{figure}

In addition to spatial filtering, the fibers of GRAVITY are used to control the dOPD between the science and phase reference objects and the polarization of the transported light (see Fig.~\ref{fig:Fibers}). The fiber differential delay lines use between 15.9\,m and 17.5\,m of fiber wrapped on two half-spools whose distance can be varied with a 100\,$\mu$m stroke piezo translation stage. The variable stretching of the fibers allows to produce variable delays up to 6\,mm. The delays are controlled with a combination of strain gauge feedback for absolute positioning with few 10\,$\mu$m accuracy and the GRAVITY metrology system for nanometer relative accuracy. 

The second kind of actuators are fiber polarization rotators in which one meter of fiber is twisted with a stepper motor to rotate the polarization and match the polarization axes for all GRAVITY baselines. The maximum stroke is slightly larger than 180$^\circ$ with an accuracy of 0.2$^\circ$. The fiber differential delay lines and the fiber polarization rotators are connected to the fiber coupler fibers and feed the fiber bundle connected to the integrated optics chip. The throughput of the whole fiber chain excluding coupling losses exceeds 87.5\,\%. The fibers and fiber control unit were developed by Le Verre Fluor\'e in collaboration with LESIA and IPAG.

\subsubsection{Integrated Optics}

\begin{figure}
\vspace{0 cm}
\centering
\includegraphics[width=\hsize]{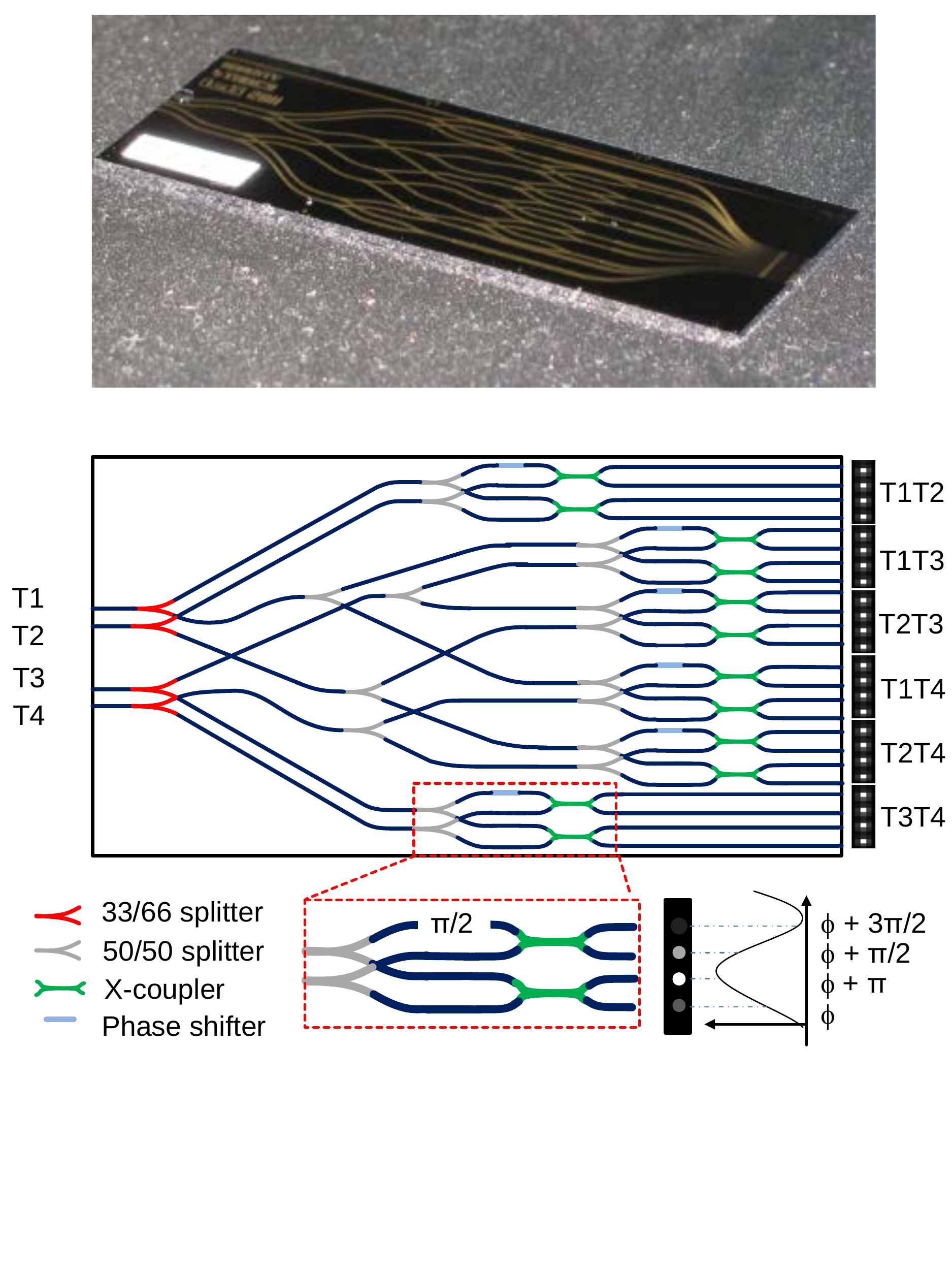}
\vspace{-2 cm}
\caption{Integrated optics beam combiner: the actual interference happens in an etched silica-on-silicon integrated optics chip. The top panel shows a photograph of the GRAVITY integrated optics, the lower panel the design of the circuit. The integrated optics contains all functions (dark blue: waveguides, red and gray: beam splitter, light blue: phase shifter, green: X coupler) for a pairwise combination of the four telescopes and fringe sampling. The light from the four telescopes T1 ... T4  is fed by single-mode fibers from the left, the output on the right is the interference of the six telescope combinations for a relative phase shift of 0$^{\circ}$, 90$^{\circ}$, 180$^{\circ}$, and 270$^{\circ}$, respectively.}
\label{fig:IntegratedOptics}
\end{figure}

The two beam combiners \citep{2014SPIE.9146E..1JJ} for the reference star and the science object are integrated optics (IO) chips -- the optical equivalent of electronic integrated circuits. These beam-combiners are directly fed by the single-mode fibers, and provide instantaneous fringe sampling for all six baselines. Fig.~\ref{fig:IntegratedOptics} shows a photograph and a schematic of the GRAVITY integrated optics. 

The fringe coding is a pair-wise simultaneous ABCD sampling, which provides in its four outputs the interference intensity at roughly $0^\circ$, $90^\circ$, $180^\circ$, and $270^\circ$ relative phase shift, respectively. As opposed to earlier implementations of the ABCD scheme (e.g., \citealt{1999ApJ...510..505C}), which apply a temporal modulation to sample the fringe, the GRAVITY beam combiner gives an instantaneous measurement of the fringe parameters. This beam combination is implemented as a double Michelson beam combiner: two successive single-mode splitters with theoretical coupling ratios of 66/33 and 50/50, respectively, split the light from each telescope in three beams with the same intensity, achromatic $\frac{\pi}{2}$ phase shifters introduce the above mentioned relative phase shifts, and two beam combiners -- X-couplers -- create the four outputs in phase quadrature. The beam combiners have therefore 24 outputs -- six baselines with four fringe samples each -- feeding the spectrometers. The integrated optics beam combiners are operated at 200K to avoid thermal background. 

The GRAVITY integrated optics beam combiners have been produced by plasma-enhanced chemical vapor deposition of phosphor-doped silica on a silicon wafer and manufactured by CEA/LETI. This technology is widely used in telecommunications up to a wavelength of 1.6\,$\mu$m and has been successfully applied in astronomical interferometry in the H-band \citep{2011A&A...535A..67L}. The GRAVITY challenge was to port this technology to longer wavelengths with a dedicated development program, first with a series of prototypes for individual functions, and then for the full integrated optics. The transmission ranges from $\approx 68$\,\% for short wavelengths around $2.0\,\mu$m to $\approx 23$\,\% at long wavelengths around $2.45\,\mu$m, with a mean a transmission of $> 54$\,\% in this wavelength range.  

\subsubsection{Spectrometer}
\label{sec:Spectrometer}

GRAVITY includes two spectrometers \citep{2014SPIE.9146E..29S} for the $1.95-2.45\,\mu$m wavelength range for the simultaneous detection of the interferometric signals of two astronomical sources: a potentially faint science object and a brighter fringe-tracking object. To minimize the thermal background, the two spectrometers are operated at 85~K. The optical input of each spectrometer consists of the 24 output channels of the respective integrated optics beam combiner. The fringe-tracking spectrometer is optimized for high readout frame rates in the kilohertz (kHz) regime at low spectral resolution with six spectral pixel, while the science spectrometer is optimized for second- to minute-long integration times and allows to select from three spectral resolutions of $R \approx 22$, $R \approx 500$ and $R \approx 4500$. Both spectrometers can be operated with or without splitting the linear polarization of the star light. The spectrometer also feeds the laser metrology backwards into the integrated optics, through which it is propagated up to the four telescopes. 

\begin{figure}
\vspace{0 cm}
\centering
\includegraphics[width=\hsize]{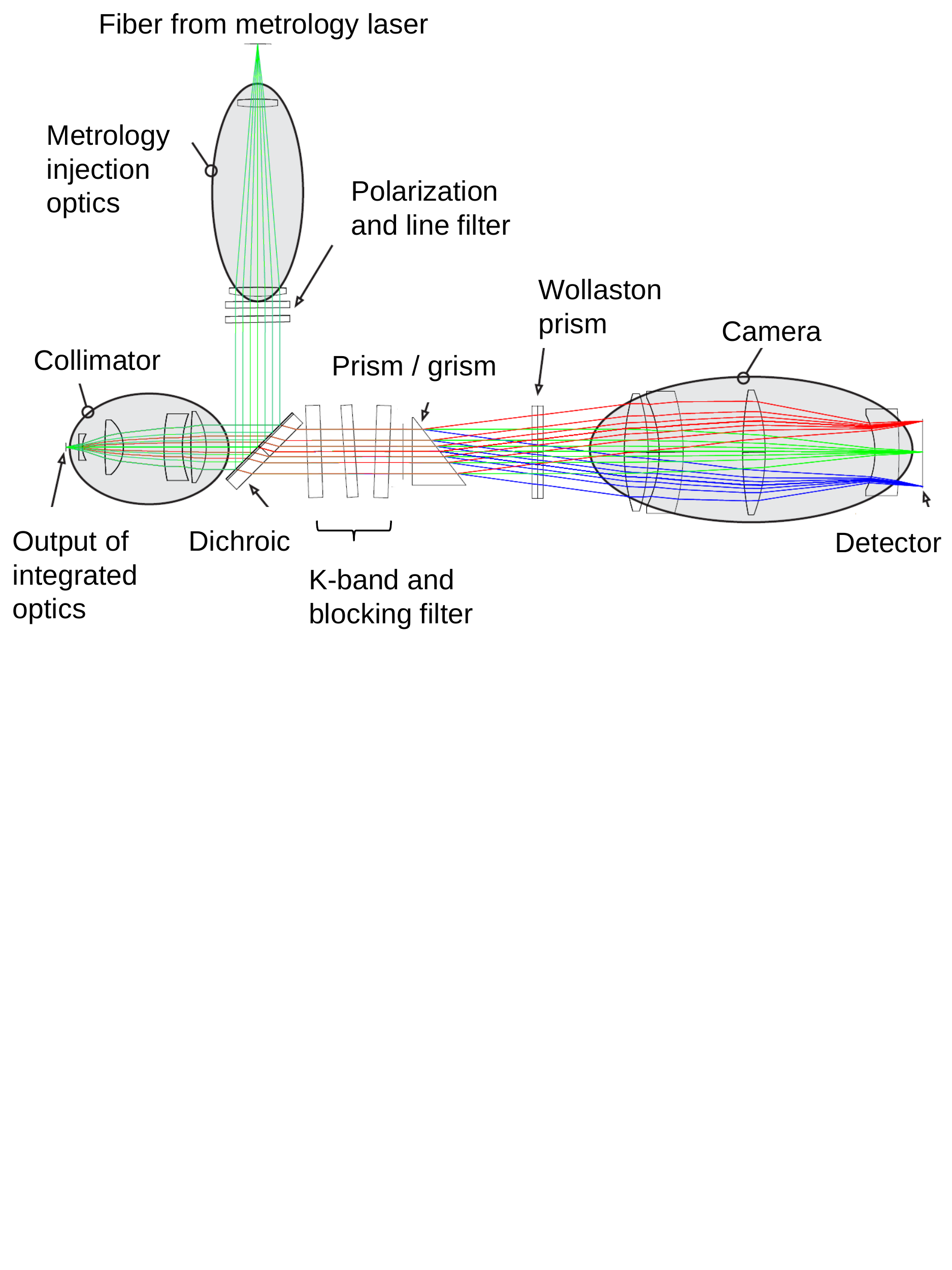}
\vspace{-5.5 cm}
\caption{Optical design of the two spectrometers of GRAVITY for the example of the science spectrometer in high spectral resolution polarimetric configuration. The dispersive elements and the Wollaston prism are mounted on cryogenic exchange mechanisms to provide user selectable configurations. The detector is mounted on cryogenic linear stages to allow  for proper focusing.}
\label{fig:Spectrometer}
\end{figure}

The optical design of both spectrometers (see Fig.~\ref{fig:Spectrometer}) is mostly identical and differs only in the dispersive elements, the angle of the Wollaston prisms, and the $F\#$ of the cameras. The beams from the 24 outputs of the integrated optics beam combiner are collimated by a four-lens system with $F\# \approx 2.27$ to a beam diameter of 24\,mm. After passing the dichroic of the laser metrology injection (the dichroic is only reflective at the laser wavelength of $1.908\,\mu$m) the light is filtered by an astronomical K-band filter and two filters to block the powerful $1.908\,\mu$m metrology laser with a total optical density of $OD \geq 16$. Following the pupil stop, the light is spectrally dispersed. In the fringe-tracking spectrometer the dispersion is achieved by a double prism made from Barium Fluoride and fused silica. In the science spectrometer the low spectral resolution uses a single fused silica prism, while the medium and high spectral resolution are produced by directly ruled grisms made from Zinc Selenide. A deployable Wollaston prism made from Magnesium Fluoride allows splitting the orthogonal directions of linear polarization. The camera optics of the science spectrometer is a four-lens system, the fringe tracker camera optics is a three-lens system including two aspherical surfaces. We chose a $F\# \approx 5.5$ for the science spectrometer to spread the high resolution spectrum over the full 2048 pixel of a HAWAII2RG detector array (see Section \ref{Detectors} on the detectors used in GRAVITY). The fringe-tracking spectrometer camera has an $F\# \approx 1.8$ to achieve an ensquared energy of $> 90$\,\% within a single $24\times24\,\mu\text{m}^2$ pixel of the SAPHIRA detector array. 

The metrology injection is identical in both spectrometers. Two metrology fibers -- enough to feed all four telescopes, because the laser light is split in the beam-combiner -- per spectrometer are each mounted on piezo-driven three-axis positioners on top of the spectrometer housing. The emerging beams are collimated by a two-lens aspheric system with $F\# \approx 4.35$ to a beam diameter of 24\,mm. After a pupil stop, the beams pass a $1.908\,\mu$m narrow-band line filter and a linear polarization filter, before they are reflected by the 45$^\circ$ dichroic, and focused by the spectrometer collimator in reverse direction on two outputs of the integrated optics beam combiner. From here the metrology laser traces exactly the stellar light path back through the integrated optics, fibers, and optics up to the telescopes.

\subsubsection{Beam combiner instrument throughput}

\begin{figure}
\vspace{0 cm}
\centering
\includegraphics[width=\hsize]{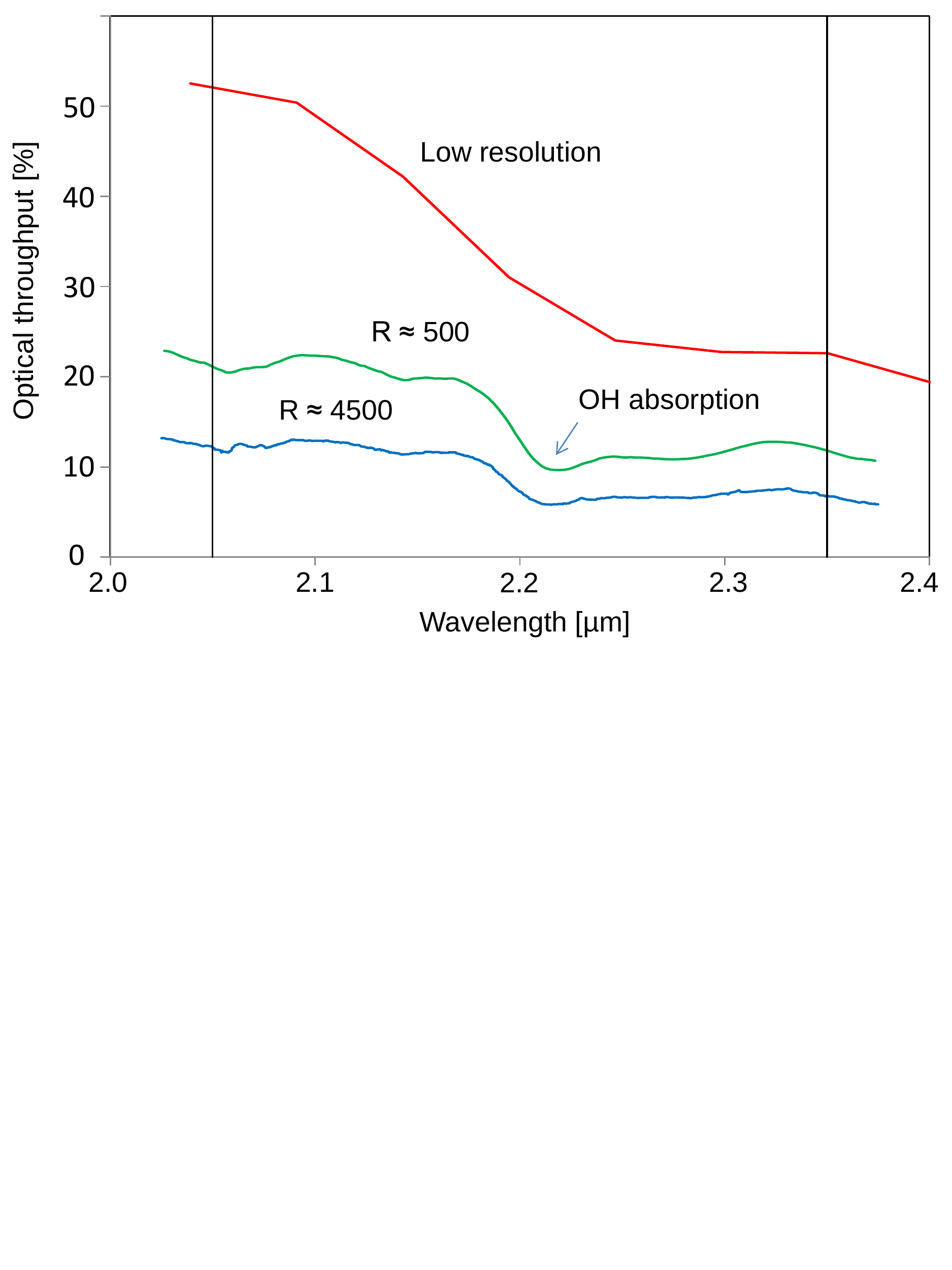}
\vspace{-6 cm}
\caption{Optical throughput of the beam combiner instrument -- excluding telescopes, VLTI beam relay, and detector quantum efficiency: low (red), $R \approx 500$ (green) and $R \approx 4500$ (blue) spectral resolution. The fringe tracker throughput is comparable to the science spectrometer at low resolution. The lower throughput of the $R \approx 500$ and $R \approx 4500$ spectral resolution is dominated by losses from the grisms. }
\label{fig:Throughput}
\end{figure}

We measured the optical throughput of the beam combiner instrument with a blackbody light source installed at the instrument input. Fig.~\ref{fig:Throughput} shows the optical throughput of the beam combiner instrument without the Wollaston prisms and excluding the detector quantum efficiencies for the different spectrometer modes. The average optical throughput of the beam combiner instrument in the $2.05-2.35\,\mu$m wavelength range is about 34\,\% when using the low spectral resolution prism. The optical throughputs with the $R \approx 500$ and $R \approx 4500$ grisms are about a factor two and four lower (indicative of their respective efficiencies), and on average about 16\,\% and 9\,\%, respectively. The prominent drop in optical throughput at a wavelength of 2.2\,$\mu$m is mostly due to OH absorption in the silica of the integrated optics \citep{2012SPIE.8445E..2XJ}. The overall quantum efficiency is typically around $0.1-1$\,\%, including the losses from telescopes and the VLTI beam relay with a K-band transmission of approximately 30\,\%, the coupling losses into the single-mode fibers from the mismatch between the uniform telescope beam and the fiber's Gaussian mode, the seeing (AT), imperfect adaptive optics correction (UT) and guiding errors resulting in a coupling efficiency of a few\,\% to a few times 10\,\%, and the detector quantum efficiency of 80\,\%.

\subsubsection{Acquisition camera and laser guiding system}
\label{AcquisitionCameraAndLaserGuidingSystem}

\begin{figure}
\vspace{0 cm}
\centering
\includegraphics[width=\hsize]{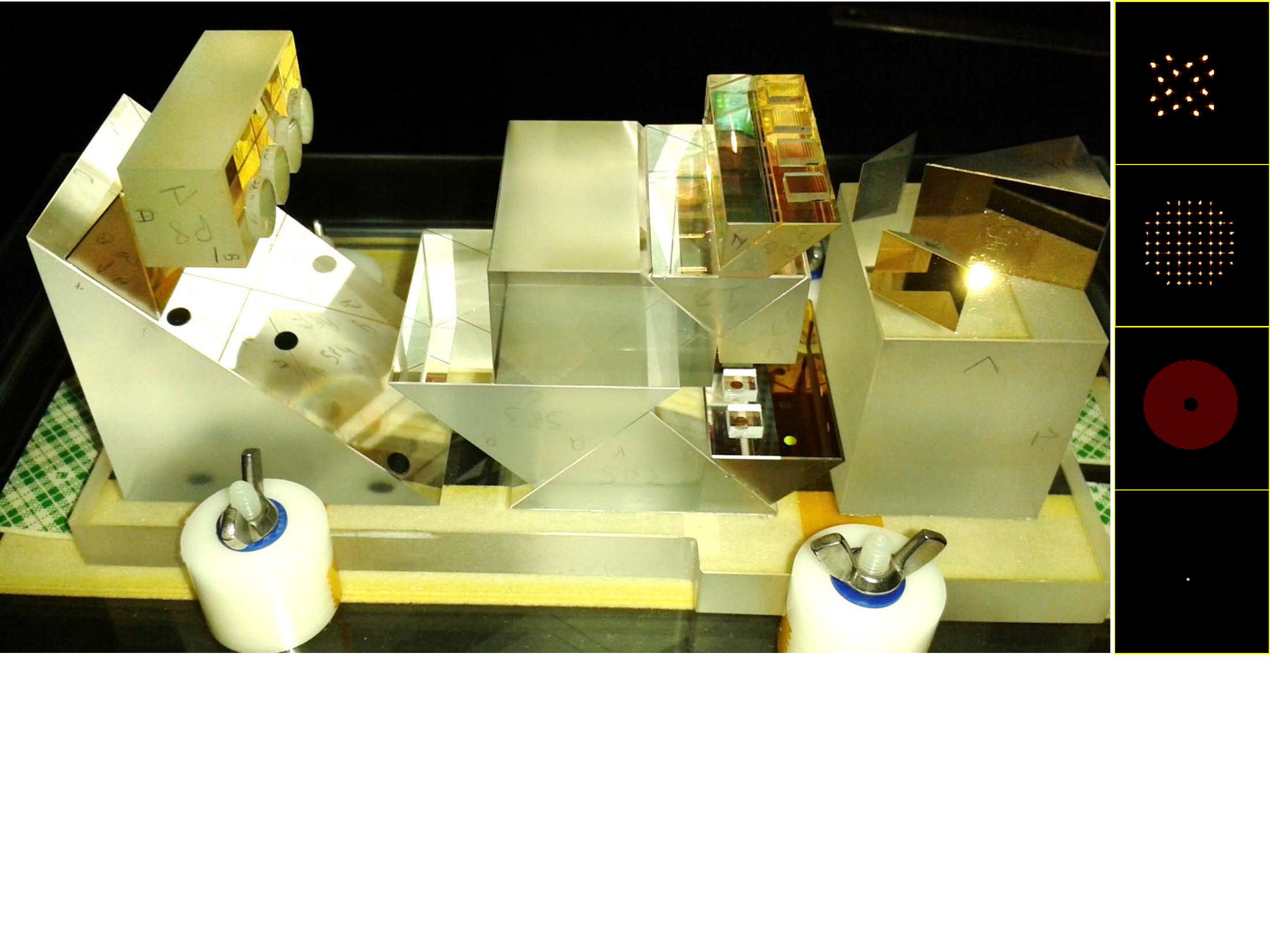}
\vspace{-2 cm}
\caption{Acquisition camera beam analyzer: the acquisition and guiding camera provides, for all four telescopes simultaneously, images of the pupil guiding lasers, a Shack-Hartman wavefront sensor, the pupil illumination, and the field (right: top to bottom, the figure shows only one of the four telescopes). All optical functions are implemented in a single, complex beam-analyzer optics (photograph), located directly in front of the detector. A dichroic beam splitter at the entrance of the beam analyzer redirects the 1200\,nm pupil guiding laser to the 2x2 lenslet. H-band light is twice split to image the field for acquisition and guiding, and to feed the Shack-Hartman sensor and pupil viewer.}
\label{fig:BeamAnalyzer}
\end{figure}

The acquisition camera \citep{2012SPIE.8445E..34A} provides simultaneously a field image, a pupil image, a Shack-Hartmann wavefront sensor image, and a pupil tracker image for all four telescopes. The H-band ($1.45-1.85\,\mu$m) field image is used for acquisition and to control low frequency image drifts. High frequency image motion from air turbulence in the optical train of the VLTI, which are not seen by the wavefront sensors located in the Coud\'e rooms of the telescopes, are measured with a 658\,nm laser beacon launched at the star-separator and detected with position sensitive diodes (Pfuhl et al. 2014). The pupil tracker is a 2$\times$2 Shack-Hartmann-like lenslet in the focal plane and is fed by the 1200\,nm pupil guiding lasers launched from the telescope secondary mirror spider arms. This pupil tracker measures both lateral and longitudinal (focus of pupil) pupil motion, and sends corresponding corrections to the instrument internal pupil actuator and the VLTI main delay line variable curvature mirror. The 9$\times$9 Shack-Hartman-sensor is used to focus the ATs, and offers the possibility to measure and correct non-common path aberrations in combination with the UT adaptive optics. All functions of the analyzer optics are implemented through fused silica micro-optics directly in front of the detector (see Fig.~\ref{fig:BeamAnalyzer}). 

\subsubsection{Laser metrology}
\label{LaserMetrology}

\begin{figure}
\vspace{0 cm}
\centering
\includegraphics[width=\hsize]{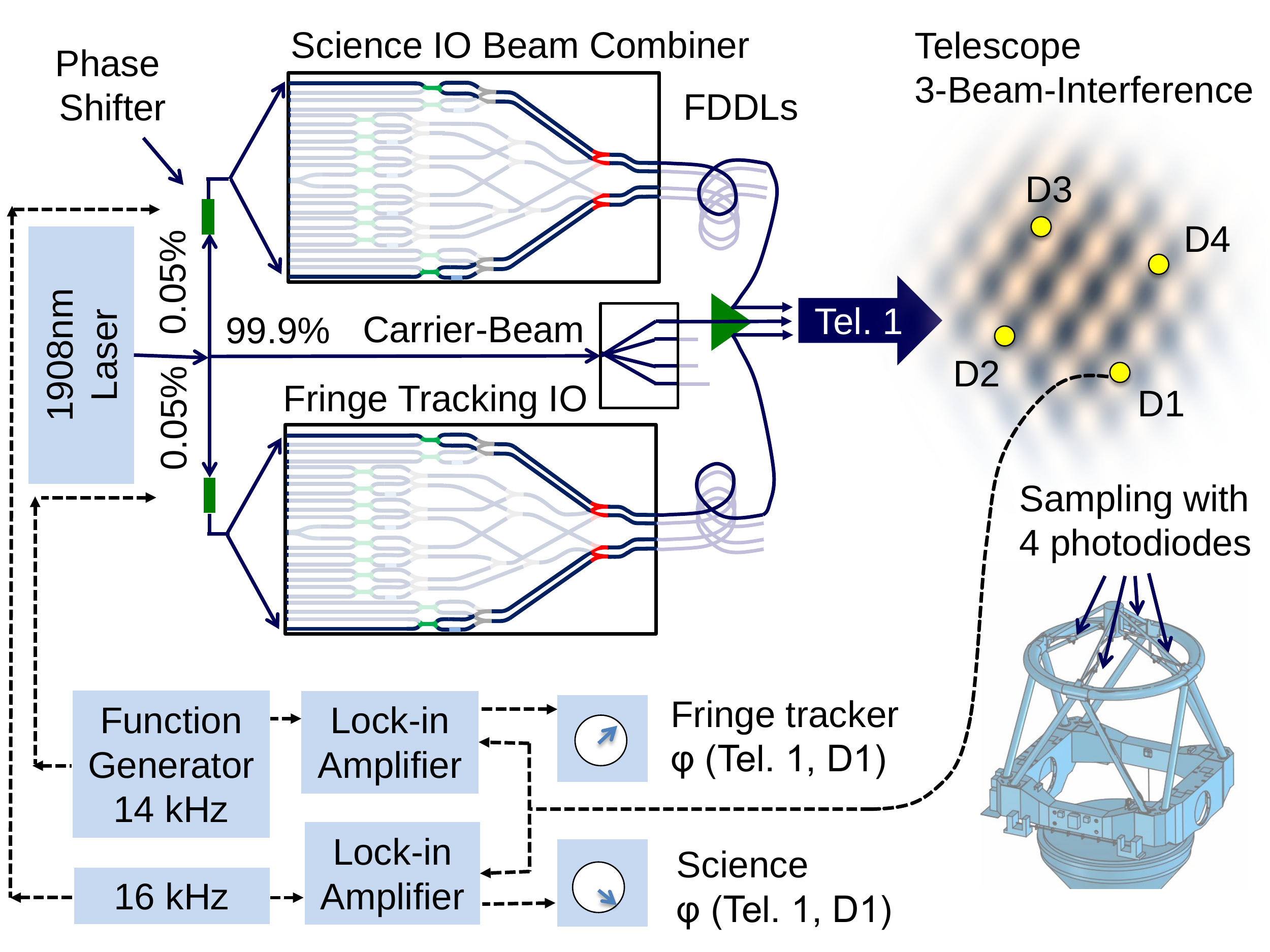}
\vspace{0 cm}
\caption{Laser metrology: the schematic shows the working principle of measuring the optical path difference between the two beam combiners and the telescope. The metrology laser is launched from the spectrometers and is detected above the primary mirror at the telescope secondary mirror spider arms. To minimize the laser power in the spectrometer and optical fibers, we do not directly measure the interference between the laser coming from the two beam combiners, but measure separately for the fringe tracker and science beam combiner the phase relative to a very bright, third ``carrier'' beam. The actual phase measurement is implemented by modulating the phase of the fringe tracker and science metrology at different frequencies, and measuring the cosine and sine components of the respective interference with the third beam through lock-in amplifiers.}
\label{fig:Metrology}
\end{figure}

The GRAVITY metrology \citep{2016SPIE.9907E..22L} measures the differential optical path between the two stars introduced by the VLTI beam relay and the beam combiner instrument. Unlike similar path length metrologies (e.g., PRIMET, \citealt{2003SPIE.4838..983L}), it does not measure the differential optical path between two telescopes for each star, but measures the differential optical path between the two stars for each telescope. This scheme has the advantage of being largely insensitive to vibrations in the VLTI optical train, and because operated in single path, eases the implementation inside the cryogenic instrument. The metrology receivers are mounted above the telescope primary mirror on the secondary mirror spider arms, which allows tracing all optical elements in the path, and even more important, to provide a physical and stable realization of the narrow angle astrometric baseline \citep{2013ApJ...764..109W, 2014A&A...567A..75L}. The metrology laser is a high-power ($\sim$1\,W), high stability ($< 30$\,MHz), linearly polarized continuous wave fiber laser with a wavelength of 1908\,nm. Two low-power beams are injected through the spectrometers into the exit of the integrated optics beam-combiners. 

The two beams are then superposed with a third beam launched in free beam from behind the Fiber Coupler, which serves as an optical amplifier. A direct detection of the interference between the fringe tracker and science metrology laser would require so high power levels in the fibers that the ineleastic backscattering from the fluorescence from Holmium and Thulium contamination and the Raman effect would completely overpower the astronomical signal at wavelengths up to about 2.15\,$\mu$m. We therefore use the interference with a much brighter third beam, thereby reducing the required flux levels in the fibers, and accordingly the backscattering, by an equivalent factor of 1000. The metrology fringe sensing is integrated as a phase-shifting interferometer with lock-in amplifier signal detection between 10\,kHz and 20\,kHz, with separate frequencies for the fringe tracker and science metrology.

\subsubsection{Detectors }
\label{Detectors}

The fringe-tracking detector is a SELEX SAPHIRA $256\times320$\,pixel, near-infrared Mercury Cadmium Telluride (HgCdTe), 2.5\,$\mu$m cutoff, electronic avalanche photo diode (eAPD) array with a pixel size of 24\,$\mu$m \citep{2016SPIE.9909E..12F}. These detectors have been developed in the context of GRAVITY in collaboration with SELEX to overcome the CMOS noise barrier, which so far was limiting the performance of near-infrared sensors at high frame rates of a few hundred Hz. The GRAVITY detectors overcome this noise barrier by avalanche amplification of the photoelectrons inside the pixel. After several development cycles, the eAPD arrays have matured and resulted in the SAPHIRA arrays as used in GRAVITY. The fringe tracker reads twenty-four $32\times3$\,pixel wide stripes at 300\,Hz -- 1\,kHz, and uses Fowler sampling with four reads at the beginning and end of the exposure. The fringe-tracking detector is run at a temperature of 95 K. The 2\,$\mu$m quantum efficiency is about 70\,\% \citep{2014SPIE.9148E..17F}. The eAPD is operated at a reverse bias voltage of 11.8 V, resulting in an eAPD gain of $\sim 36$. The resulting effective read noise is $< 1$\,e$^-$ rms, and the excess noise from the amplification process is 1.3. 

The science spectrometer and the acquisition camera are each equipped with a TELEDYNE $2048\times2048$\,pixel, 2.5\,$\mu$m cutoff wavelength HgCdTe, 18\,$\mu$m pixel size, HAWAII2RG detector \citep{2008SPIE.7021E..0PF}. The detectors are operated in non-destructive -- sampling up the ramp -- read mode, using the 32 100\,kHz analog outputs. The quantum efficiency at a wavelength of 2\,$\mu$m is around 80\,\%. The correlated double sampling read noise of the acquisition camera and science spectrometer detectors is 13\,e$^-$ and 12\,e$^-$ rms, respectively. The effective read noise for sampling with 32 Fowler pairs is about 3\,e$^-$ rms. All detectors of GRAVITY are controlled with the ESO NGC detector electronics \citep{2009Msngr.136...20B}.

\subsubsection{Calibration unit}

The calibration unit \citep{2014SPIE.9146E..1UB} provides all functions to test and calibrate the beam combiner instrument. It is directly attached to the beam combiner instrument in front of the cryostat and simulates the light from two stars and four telescopes. The artificial stars are fed by halogen lamps or an Argon spectral calibration lamp. The calibration unit further provides four motorized delay lines to co-phase the beams, metrology pickup diodes to simulate astrometric observations, linear polarizing filters to align the fiber polarization, tip/tilt and pupil laser beacons for testing the pupil tracker and fast guiding, and rotating phase screens for simulating seeing residuals.

\subsubsection{Instrument control software and hardware}

The instrument software \citep{2014SPIE.9146E..2AO, 2014SPIE.9146E..2BB} is implemented within the ESO control software framework \citep{2008SPIE.7019E..0QP}. In addition to the basic Instrument Control Software (ICS), which handles motors, shutters, lamps, etc., the instrument software also includes the Detector Control Software (DCS, \citealt{2006ASSL..336..589C}), several special devices, field bus devices \citep{2014SPIE.9152E..07K}, and various real time algorithms. The latter are implemented using the ESO Tools for Advanced Control (TAC, \citealt{2004SPIE.5496..155B}) and run at a frequency of up to 3.3\,kHz. In total, the instrument has more than a hundred devices. 

The various control and data acquisition processes are distributed over a total of six Linux workstations (instrument workstation, three detector workstations, workstation for analyzing the fringe tracker residuals and updating the Kalman model, and data recorder workstation),  seven VxWorks computers for controlling and commanding the various hardware functions (e.g., motors, piezos, shutters, lamps, and lasers) and real-time applications (metrology, phase sensor, OPD controller, differential delay line controller, and tip/tilt/piston controller), partly connected through a reflective memory ring, and two programmable logic controllers from Beckhoff (stepper motor control) and Siemens (cryo- and vacuum control).

\subsubsection{Fringe-tracking}
\label{FringeTracking}

\begin{figure}
\vspace{0 cm}
\centering
\includegraphics[width=\hsize]{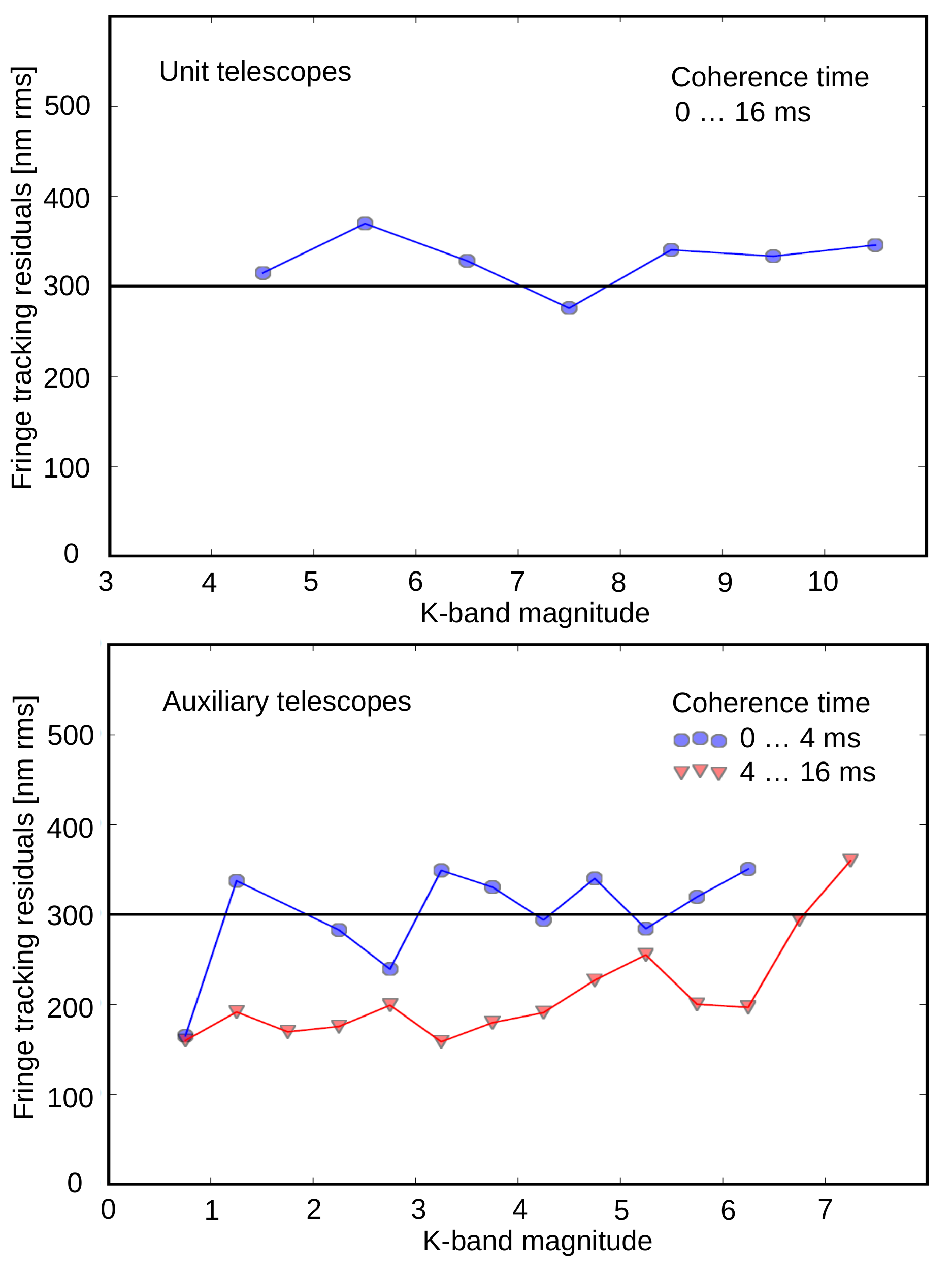}
\vspace{0 cm}
\caption{Fringe-tracking performance: the figures show the fringe-tracking residuals as a function of the reference stars's K-band correlated magnitude for UTs (top) and ATs (bottom), respectively. The horizontal lines indicate residuals of 300\,nm rms, for which the fringe contrast in long exposures is reduced by $\sim 30$\,\% in the K-band. For the ATs, we plot the fringe-tracking residuals separately for good seeing with long atmospheric coherence times $\tau_0 > 4$\,ms (red), and for short coherence times $\tau_0 < 4$\,ms (blue). We do not have enough statistics to make this distinction for the UTs. The OPD residuals for long coherence times are typically 200\,nm and 300\,nm rms, the limiting magnitudes around $m_K \approx 7$\,mag and $m_K \approx 10$\,mag for the ATs and UTs, respectively.}
\label{fig:FringeTracker}
\end{figure}

The fringe-tracking system \citep{2012A&A...541A..81M, 2014A&A...569A...2C} stabilizes the fringes to allow for long exposures with the science spectrometer, and to operate the instrument close to the white-light condition for accurate phase measurements when doing astrometry. A cascade of real-time and workstation computers \citep{2016SPIE.9907E..21A}, connected via a reflective memory ring, analyzes the detector images arriving at 300\,Hz to 1\,kHz, runs the actual control algorithm, applies it to the actuators, and optimizes the control parameters at runtime. 

The first module -- the fringe sensor -- receives the data stream from the fringe-tracking detector and computes the phase delay and group delay for each of the six baselines. The second module -- the OPD controller -- then calculates the correction signal for the piston actuators. It also hosts the state machine to switch between fringe search, group delay tracking to center the fringe, and phase-tracking to stabilize the fringe. The fringe tracker is based on a Kalman controller for optimum correction of the atmospheric and vibration-induced piston and to mitigate flux dropouts from the fluctuating injection in the fibers. The Kalman model for the piston is an autoregressive model of order 30, the system- and piezo-response is modeled with a fourth order autoregressive model. The Kalman model parameters are automatically updated every few seconds by a non-realtime workstation analyzing the actual OPD residuals and actuator commands. The piston commands from the fringe tracker are finally merged with the laser guiding measurements in a third module -- the tip/tilt/piston controller -- and applied to the instrument internal tip/tilt/piston piezo actuator. Low frequency OPD variations are offloaded to the VLTI main delay line. The fringe tracker is synchronized with the science detector to sample the science fringes at discrete phase offsets, and to guarantee that $2\pi$ phase corrections -- resulting from the drift between phase and group delay -- are only applied between exposures. 

The fringe-tracking performance depends on the atmospheric coherence time $\tau_0$ and the K-band correlated magnitude of the reference star. Fig.~\ref{fig:FringeTracker} shows this dependence for the case of on-axis observations, for which the light is equally split between the fringe tracker and science spectrometer. To increase the statistics -- especially for the comparably few UT commissioning observations -- we have also included off-axis observations by subtracting $\sim 0.75$\,mag from the stars apparent magnitude. For good observing conditions with an atmospheric coherence time $\tau_0 > 4$\,ms, the fringe-tracking limiting magnitude is around $m_K \approx 7$\,mag for the ATs and $m_K \approx 10$\,mag for the UTs,  respectively. The limiting magnitudes for off-axis observations, for which all the light from the reference stars is used for fringe-tracking, are approximately 0.75\,mag fainter. The OPD residuals -- calculated over a time window of typically few minutes -- are around 200\,nm and 300\,nm rms, respectively. The larger fringe-tracking residuals for the UTs are caused by uncorrected vibrations of the telescopes and Coud\'e optics.

\subsubsection{Differential delay control}
\label{sec:DifferentialDelayControl}

The dOPD between the fringe tracker and science objects is continuously compensated by the instrument internal fiber differential delay lines (Section \ref{SingleModeFibersAndFiberControl}). The differential delay lines are preset using the strain gauge feedback from the piezo actuators, the stabilization is done on the metrology feedback. The dOPD trajectory is calculated in realtime from the object coordinates and telescope locations. The typical preset accuracy on the strain gauge is about several 10\,$\mu$m, the closed loop residuals on the metrology feedback are on nanometer level. The reflective memory ring is used to synchronize the data between the metrology realtime computer and the fiber differential delay controller. 

\subsubsection{Science fringe centering}

The uncertainty in the relative position between the fringe tracker star and the science object, as well as the hysteresis of the fiber delay line and its strain gauge feedback, limits the accuracy of the science fringe centering. The instrument software thus provides the possibility to automatically center the science fringes. This is done in a similar way as with the fringe tracker, but here analyzing the long exposures from the science spectrometer, and commanding the differential delay lines.

\subsubsection{Field stabilization}
\label{sec:FieldStabilization}

The field stabilization is implemented in a three stage control. The atmospheric- and wind-shake-induced image motion is corrected at the telescope level, in the case of the UTs with the GRAVITY Coud\'e Infrared Adaptive Optics (Section \ref{AdaptiveOptics}) or the Multi-Application Curvature Adaptive Optics (MACAO, \citealt{2003SPIE.4839..174A}), and for the ATs with the System for Tip-tilt Removal with Avalanche Photodiodes (STRAP, \citealt{1997SPIE.3126..580B}) and in the future with the New Adaptive Optics Module for Interferometry (NAOMI, \citealt{2016SPIE.9907E..20G}). When observing with the UTs, the GRAVITY laser guiding system (Section \ref{AcquisitionCameraAndLaserGuidingSystem}) measures the image jitter between the telescope and the beam combiner instrument. This control runs at 3.3\,kHz loop rate on a real-time computer, and directly actuates the instrument internal tip/tilt/piston actuator in the fiber coupler (Section \ref{FiberCoupler}). The low frequency image drifts are measured by the acquisition and guiding camera with a frame rate of 0.75\,Hz. The image analysis \citep{2014SPIE.9146E..2CA} and the control are implemented on the instrument workstation. The typical residuals of the acquisition camera guiding are $< 0.5$\,pixel (one axis rms), corresponding to $< 0.2$\,times the H-band diffraction limit of the telescopes, or $< 9$\,milliarcsecond (mas) (UT) and $< 40$\,mas (AT).

\subsubsection{Pupil control}
\label{sec:PupilControl}

Accurate lateral and longitudinal pupil control is a prerequisite for narrow angle astrometry \citep{2014A&A...567A..75L}. The pupil position is traced by the laser beacons launched at the telescope spider arms and measured with the acquisition and guiding camera (Section \ref{AcquisitionCameraAndLaserGuidingSystem}). The image processing \citep{2014SPIE.9146E..2CA} and the control loop are implemented on the instrument workstation. The frame rate is 0.75\,Hz. The typical accuracy of the lateral pupil guiding is 0.1\,\% (one axis rms) of the pupil diameter, and about 30\,mm rms (in the 80\,mm diameter collimated beam of the VLTI main delay lines) for the longitudinal pupil guiding.  

\subsection{Adaptive optics}
\label{AdaptiveOptics}

The GRAVITY Coud\'e Infrared Adaptive Optics (CIAO, \mbox{\citealt{2016SPIE.9909E..2LS}}) is a single conjugated adaptive optics, combining a Shack-Hartmann type wavefront sensor sensitive in the near-infrared H+K-bands ($1.30-2.45\,\mu$m) with the ESO Standard Platform for Adaptive optics Real Time Applications (SPARTA, \citealt{2006SPIE.6272E..10F}), and the bimorph deformable mirror from the visual light Multi-Application Curvature Adaptive Optics (MACAO, \citealt{2003SPIE.4839..174A}). 

\subsubsection{Wavefront sensor}

\begin{figure}
\vspace{0 cm}
\includegraphics[width=\hsize]{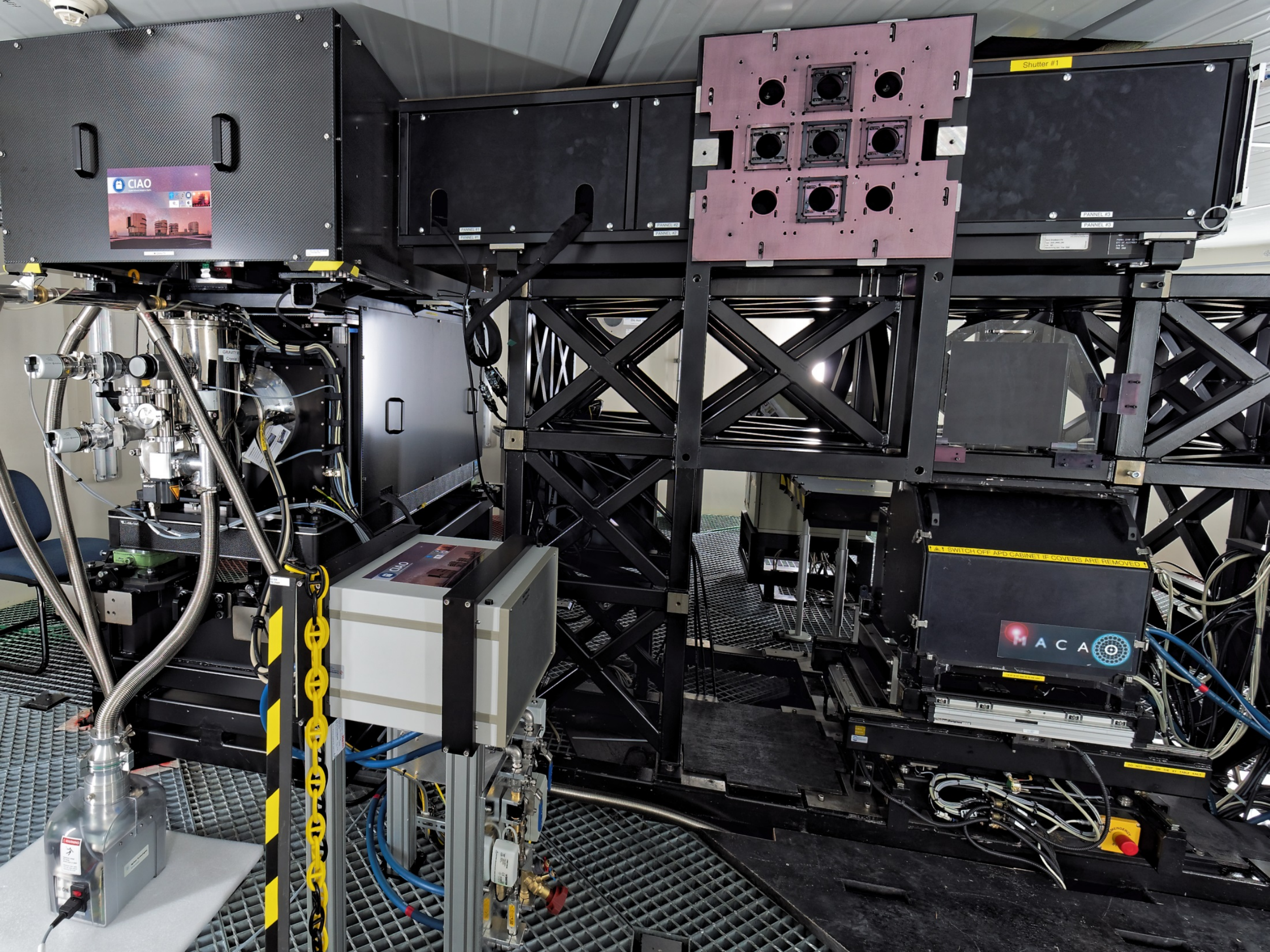}
\vspace{0 cm}
\caption{Coud\'e Infrared Adaptive Optics: the photograph shows the Coud\'e room of UT4 in June 2016. The big central structure is the star separator. The CIAO wavefront sensor comprises the tower structure on the left side, with the cryostat connected to a pump on the floor, and the readout electronics located next to it (light gray box connected to blue cooling pipes). The visual wavefront sensor MACAO is located at the lower right.}
\label{fig:CIAO}
\end{figure}

\begin{figure}
\vspace{0 cm}
\includegraphics[width=\hsize]{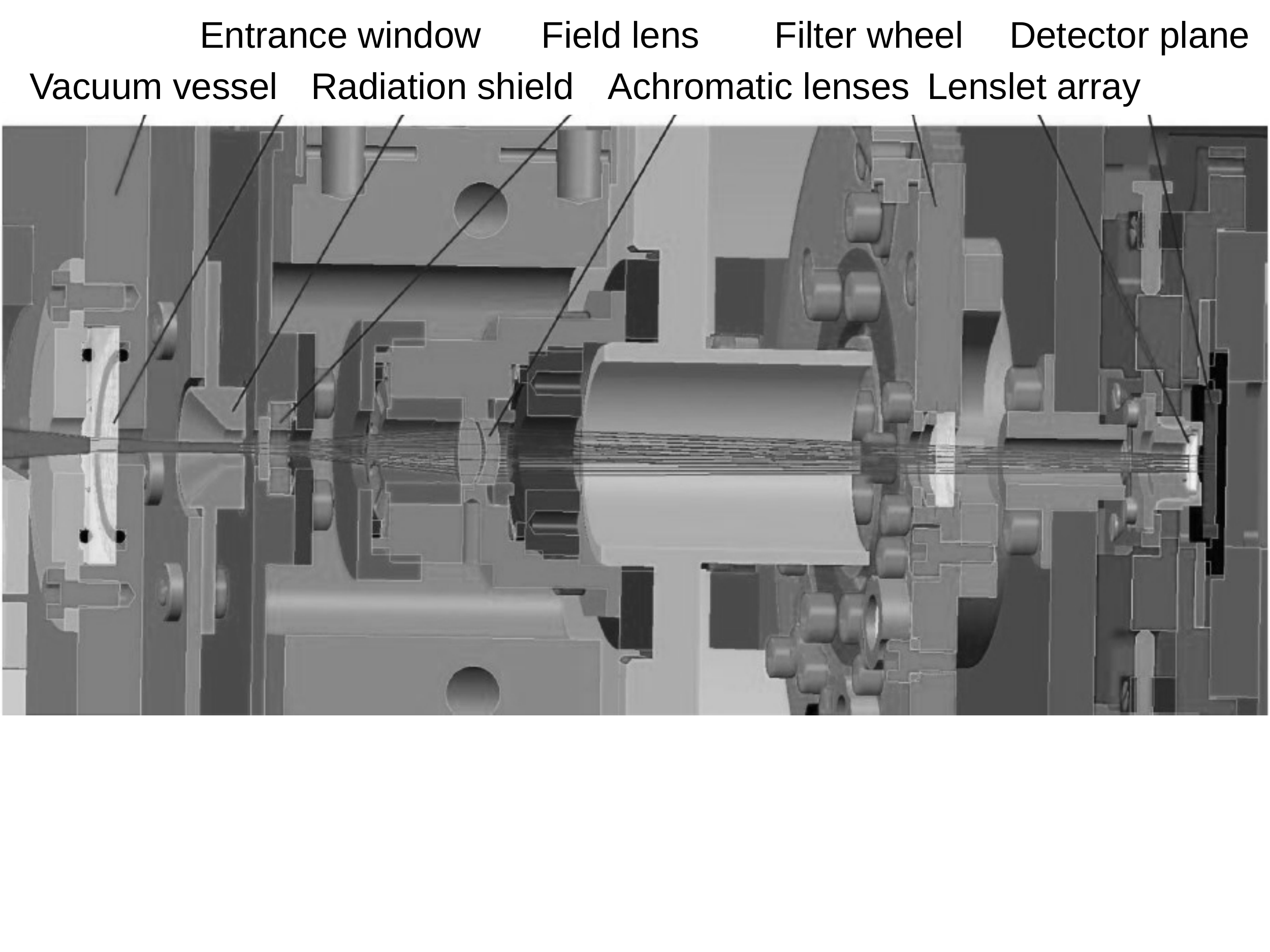}
\vspace{-1.5 cm}
\caption{CIAO cryostat: the 3D CAD drawing depicts the CIAO cryostat with its cold optics and actuators. The movable field lens, which is located behind the entrance window, controls the lateral pupil position \citep{2017JMOp...64..127D} as seen by the lenslet array. The achromatic lenses image the deformable mirror (pupil) onto the lenslet array. The filter wheel houses an H+K-band filter, two neutral density filters, a closed position, and an open position.}
\label{fig:CIAO_cryostat}
\end{figure}

The CIAO wavefront sensor has 68 active subpupils, with each subpupil corresponding to an area with a diameter of 0.9\,m projected on the UT primary mirror. The image scale on the detector is 0.5\arcsec/pixel. The instantaneous, unvignetted field of view of the wavefront sensor spans 2\arcsec, corresponding to $4 \times 4$\,pixel on the detector. To minimize crosstalk between the subpupils, each subpupil is re-imaged on an $8 \times 8$\,pixel area of the detector. The wavefront sensors use SAPHIRA eAPD detector arrays, which are the same type as used in the fringe tracker, and are described in Section \ref{Detectors}. 

The CIAO wavefront sensors are located in the Coud\'e room below each of the UTs (see Fig.~\ref{fig:CIAO}) behind the star separator \citep{2004SPIE.5491.1528D}. The star separator has access to a field of view on sky with a radius of 60\arcsec. Within this large field of view, the star separator can select and track two separate beams with a field of view of 2\arcsec\ each. The star separator also provides actuators for pupil positioning and stabilization. CIAO can pick either of the two star separator outputs, using mirrors to take advantage of all of the light when operated off-axis, or inserting a beam splitter when used on-axis together with the GRAVITY beam combiner instrument. The CIAO beam selector is prepared to host one more beam splitter in preparation for a potential upgrade for use with the future VLTI instrument MATISSE \citep{2014Msngr.157....5L}. 

In order to block parasitic light from the 1200\,nm GRAVITY pupil beacons and the 1908\,nm metrology, the CIAO cryostat  (see Fig.~\ref{fig:CIAO_cryostat}) entrance window only transmits light in the wavelength range from $1.3-2.45\,\mu$m and includes a notch filter blocking light at wavelengths between 1.85\,$\mu$m and 2.0\,$\mu$m \citep{2013OExpr..21.9069Y}.

\subsubsection{Target acquisition}

For the target acquisition, CIAO scans a field of view of typically 10\arcsec$\times$10\arcsec, automatically identifies the brightest source in this field, and selects the loop frequency and detector gain for optimal performance. The loop frequencies range from 100\,Hz (faint star case) to 500\,Hz (bright star case). The scanned field of view is displayed to verify the selected reference star, or to manually pick another source, e.g., in the case of a crowded field. After closing the adaptive optics loop, CIAO optimizes and stabilizes the pupil alignment using both its internal actuator and the star separator pupil actuator. Once the CIAO acquisition process has been completed, the control is yielded back to the beam combiner instrument for the acquisition of the interferometric targets. 

\subsubsection{Adaptive optics performance}

\begin{figure}
\vspace{0 cm}
\includegraphics[width=\hsize]{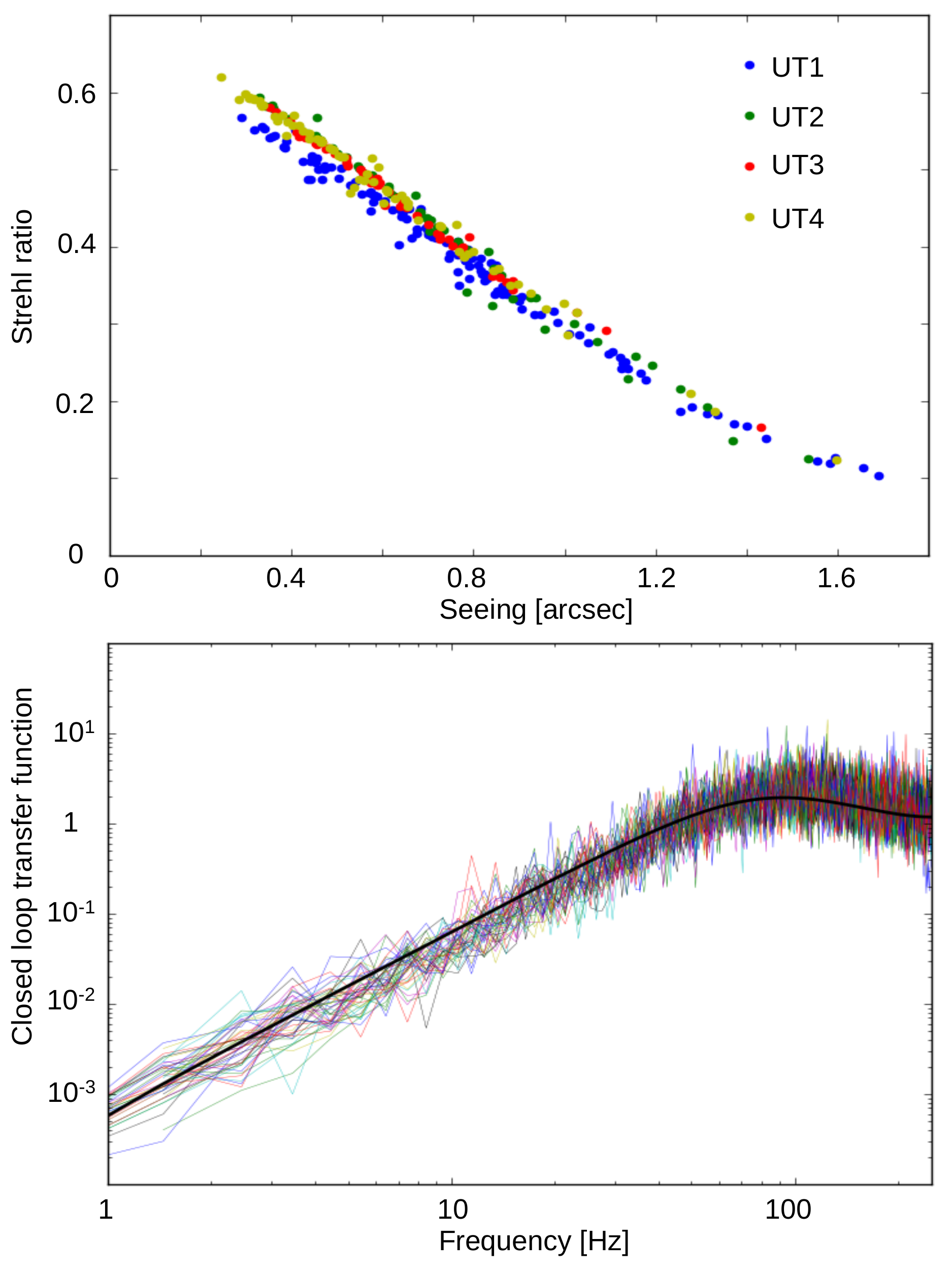}
\vspace{0 cm}
\caption{Adaptive optics performance: the top panel shows the K-band Strehl ratio as a function of atmospheric seeing (at 500\,nm). The Strehl ratio and seeing are derived from the wavefront residuals as observed with the wavefront sensor, and calibrated with K-band observations of the components of wide binary stars using the VLTI IRIS instrument \citep{2004SPIE.5491..944G}. The bottom panel shows the closed loop transfer function for Zernike modes up to order 44 as measured on the $m_K = 6.5$\,mag star GC\,IRS\,7. Both curves are representative for observations of bright objects with $m_K \lesssim 7$\,mag.}
\label{fig:CIAO_Performance}
\end{figure}

For bright sources, the adaptive optics performance is limited by the number of actuators (60) of the MACAO deformable mirror. Fig.~\ref{fig:CIAO_Performance} (top) shows the typical K-band Strehl ratio as a function of atmospheric seeing. The Strehl ratio delivered to the VLTI at a seeing of 0.7\arcsec\ and using a wavefront reference star at a separation of 6\arcsec\ is 40\,\%. CIAO fulfills or outperforms all its top-level requirements \citep{2016SPIE.9909E..2MD}, in particular for the on-axis K-band Strehl ratio of $\ge 35$\,\% and $\ge10$\,\% on stars with $m_K = 7$\,mag and $m_K = 10$\,mag, respectively.\footnote{At a zenith distance of 30$^\circ$ for the standard Paranal atmosphere with zenith seeing of 0.85\arcsec\ (corresponding to r$_0 = 0.12$\,m), coherence time of 3\,ms, and (zenith) isoplanatic angle of $\sim 2$\arcsec\ at a reference wavelength of 500\,nm.} Fig.~\ref{fig:CIAO_Performance} (bottom) shows the closed loop transfer function obtained on GC\,IRS\,7, the wavefront reference star for the observations of the Galactic Center (Section \ref{GRAVITYObservationsOfTheGalacticCenter}). In better than average observing conditions CIAO works on guide stars as faint as $m_K \approx 11$\,mag. It provides better performance than the visible MACAO on objects with $V-K \ge 4.5$\,mag. CIAO also includes neutral density filters for observations of bright stars.

\subsection{Data reduction software}
\label{DataReductionSoftware}

The data reduction software provides all routines for the calibration and reduction of the data collected with the instrument. It covers routines for the instrument calibration, single beam observations, and dual beam observations. The inputs are the science combiner detector frames, the fringe tracker frames, the metrology signals from the diodes in the fiber coupler and at the telescopes, and the images from the acquisition camera. The outputs of the data reduction software are calibrated complex visibilities, reconstructed quick-look images, and astrometry data. 

The main data reduction algorithm of the GRAVITY pipeline is based on the principle of the Pixel to Visibility Matrix (P2VM, \citealt{2007A&A...464...29T}). The P2VM characterizes the photometry, coherence and phase relations between the four inputs of the integrated optics components and their 24 outputs (six baselines times four outputs). An overall description of the algorithms used in GRAVITY is given in \cite{2014SPIE.9146E..2DL}. A complete instrument calibration data set includes the P2VMs of the two beam combiners, the map of the spectral profiles of the science spectrometer, the wavelength calibration of the fringe tracker and science spectrometer, as well as dark frames and bad pixel maps. These calibration data are computed from a series of raw files collected using the calibration unit. The wavelength scale is also derived from this sequence, using the metrology laser wavelength as the fiducial reference. The properties of the interference fringes (photometric spectra, complex visibilities, closure quantities) as a function of wavelength are computed separately for the fringe tracker and science beam combiners. In addition, when used in dual field mode, GRAVITY provides the phase of the science fringes referenced to the fringe tracker, which can be translated into an astrometric separation vector. Optionally, the pipeline can also analyze the acquisition camera frames. 

The data reduction software code is written in standard ANSI C using ESO's Common Pipeline Library \citep{2004SPIE.5493..444M}. It is made available through ESO. The MiRA image reconstruction algorithm \citep{2008SPIE.7013E..1IT, 2013EAS....59..157T} is interfaced with the GRAVITY pipeline through a dedicated processing recipe and is included in the data reduction software distribution. The GRAVITY pipeline can be executed using the {\tt esorex} command line tool, from the {\tt Gasgano} graphical user interface \citep{2012ascl.soft10020E}, or from a {\tt reflex} graphical workflow \citep{2014ASPC..485...11B}. Alternatively, a set of python routines developed by the GRAVITY consortium {\tt python\_tools} can be used to run the reduction, calibrate and visualize the raw and processed data. 

The output files produced by the data reduction software follow the Optical Interferometry FITS version 2 standard \citep{2016SPIE.9907E..10D}. They can therefore be visualized and analyzed using standard interferometric software packages such as offered by the Jean-Marie Mariotti Center, with interferometric image reconstruction codes (see, e.g., \citealt{2014SPIE.9146E..1QM} for a review of existing codes), and special analysis software as, e.g., the companion search tool CANDID \citep{2015A&A...579A..68G}.

\subsection{Measurement precision}

\begin{figure}
\vspace{0 cm}
\centering
\includegraphics[width=\hsize]{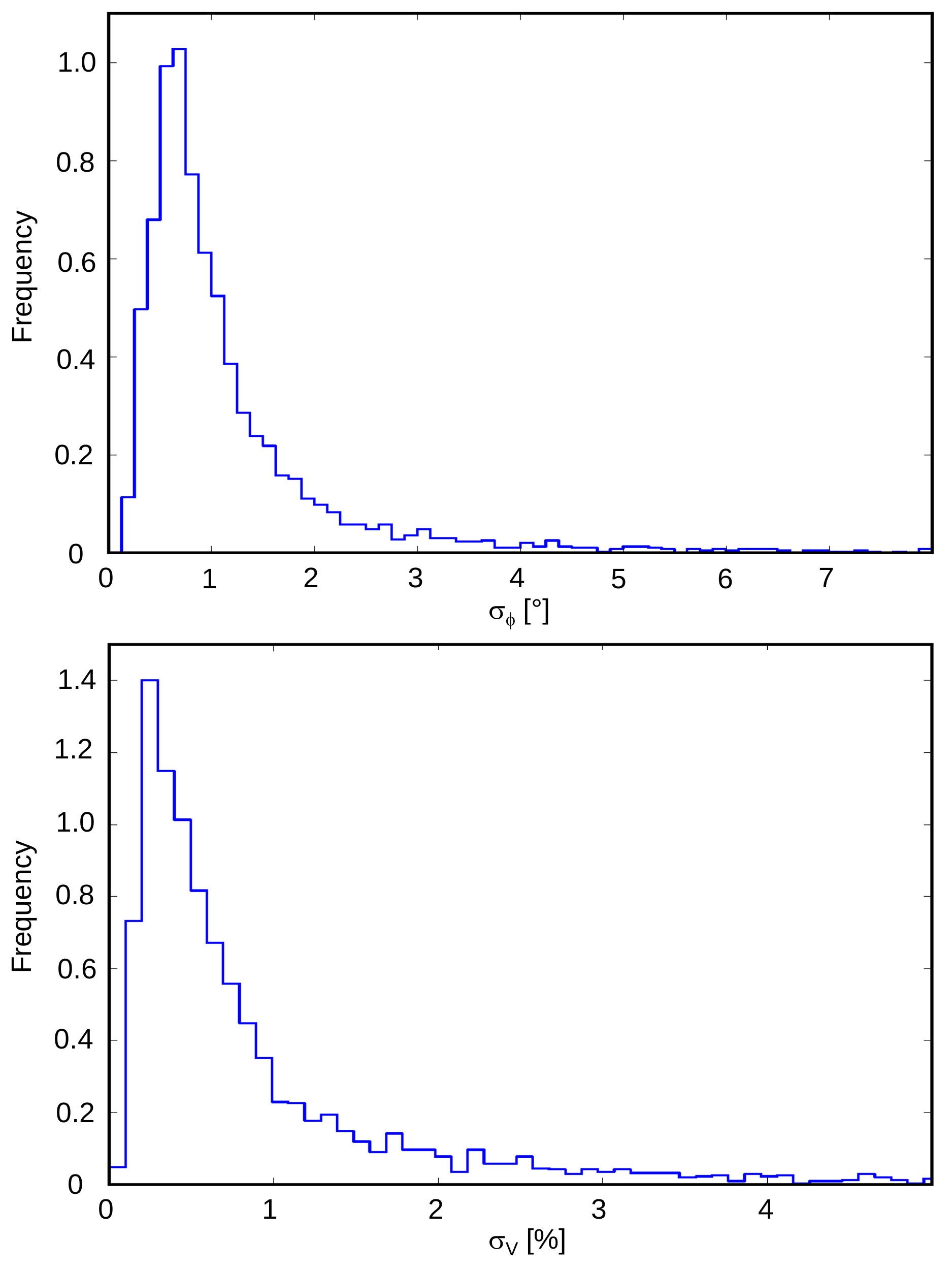}
\vspace{0 cm}
\caption{Measurement precision: the figures show the histograms of the wavelength-averaged standard deviations of the visibility phase $\sigma_\Phi$ (top) and amplitude $\sigma_V$ (bottom) for the observation of bright calibrator stars. They typically peak around $1^\circ$ and 0.5\,\%, respectively.}
\label{fig:Precision}
\end{figure}

The precision of the interferometric phase and visibility amplitude in long science exposures is a function of the source brightness and the fringe-tracking residuals. The standard deviations of these quantities are calculated by the data reduction software (see Section \ref{DataReductionSoftware}) by bootstrapping\footnote{See, e.g., Section 15.6 of \cite{2002nrca.book.....P} for an introduction to the bootstrap method.} the measurements from the individual exposures of each data set. Fig.~\ref{fig:Precision} shows the histograms of the standard deviations of the visibility phase and amplitude for bright calibrator stars observed with the ATs in good conditions, with fringe-tracking residuals smaller than 300\,nm rms. The fringe-tracker was run at a frame rate of 909\,Hz and a closed-loop cutoff frequency of around 60\,Hz \citep{2016SPIE.9907E..21A}. The data represented in the histograms are the wavelength-averaged uncertainties for each baseline and data set. We analyzed around 600 data sets. Each data set typically contains 30 exposures with individual integration times between $0.3 - 30$\,s. The precision of the visibility amplitude is about 0.5\,\%, and about $1^\circ$ for the visibility phase. The results for the UTs are similar to the ATs.


\section{First GRAVITY observations}
\label{sec:FirstGRAVITYobservations}

This section illustrates the observing modes and demonstrates the performance of GRAVITY for several archetypical objects observed during commissioning and early guaranteed time observations.

\subsection{High accuracy visibility observations of resolved stars}
\label{HighAccuracyVisibilityObservationsOfResolvedStars}

The fidelity of interferometric imaging relies on high accuracy visibilities and phases. The dynamic range -- the intensity ratio of the brightest and faintest objects detectable in the image -- is to first order inversely proportional to the noise in the visibility and phases. Also imaging resolved stars with their low contrast surface features requires a very high visibility accuracy (e.g., \citealt{2009A&A...508..923H}). We demonstrate the exquisite accuracy of GRAVITY and its visibility calibration with two examples among the best we have obtained so far, the observations of $\xi$\,Tel (Section \ref{sec:XiTel}) and 24\,Cap (Section \ref{sec:24Cap}). They are examples for the two regimes of high and very low visibilities, respectively. We took the data in medium spectral resolution ($R \approx 500$) on the nights of 7 and 8 October 2016 with the ATs in the A0-G1-J2-K0 configuration. 

\subsubsection{Limb darkening in $\xi$\,Tel}   
\label{sec:XiTel}

\begin{figure}
\vspace{0 cm}
\centering
\includegraphics[width=\hsize]{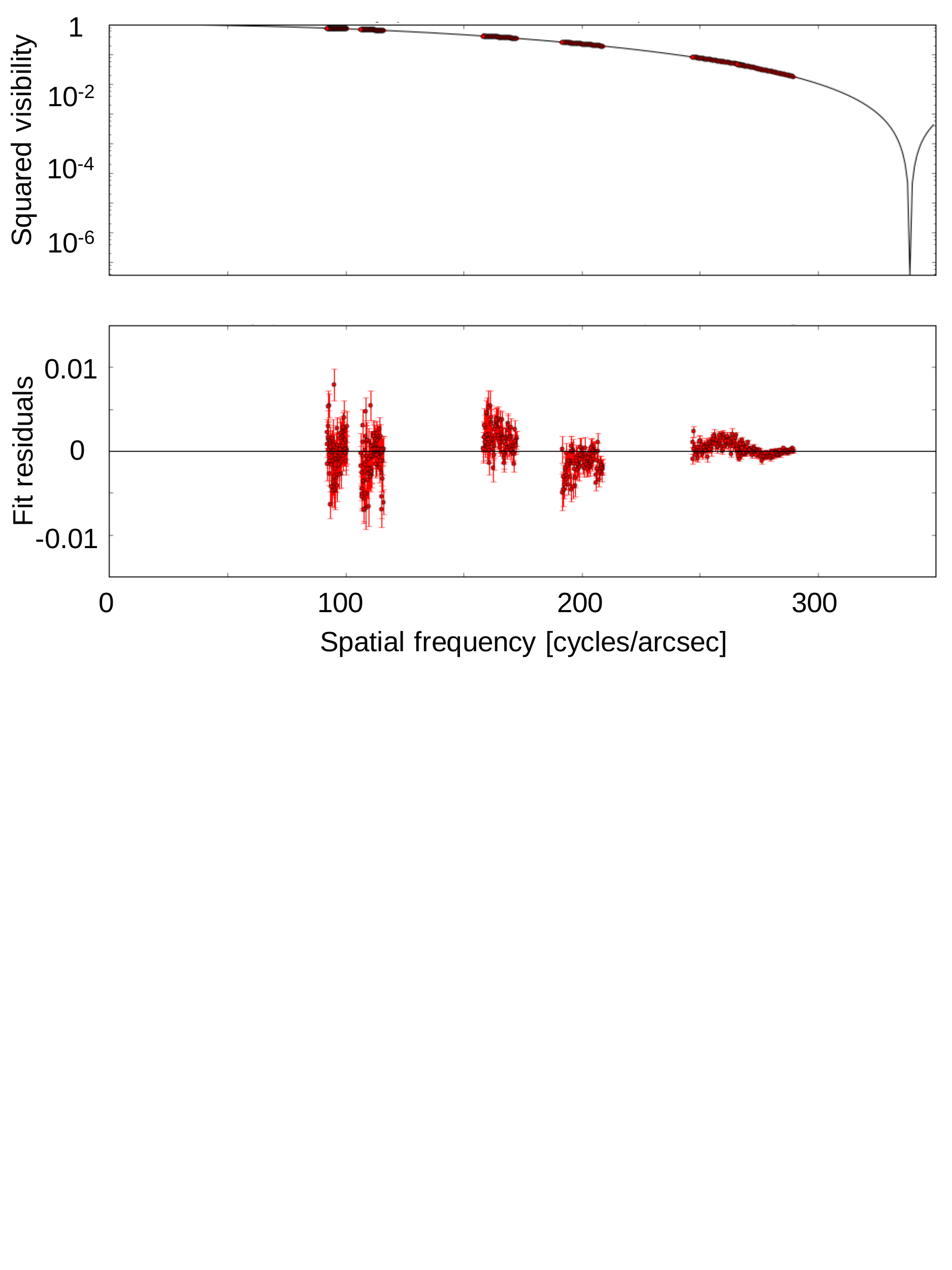}
\vspace{-6 cm}
\caption{Limb darkening in the K5 III giant $\xi$\,Tel: The top panel shows the observed squared visibility for $\xi$\,Tel. The solid line is the best fit limb darkening disc model. The fit residuals (bottom) are typically smaller than 0.005, corresponding to a visibility accuracy of better than 0.25\,\%.}
\label{fig:xiTel}
\end{figure}

We measured the accuracy of GRAVITY at high visibilities with the K5 III giant $\xi$\,Tel, a bright $m_K = 0.91$\,mag star (Fig.~\ref{fig:xiTel}). Each calibrated point is the combination of 30 exposures of 5\,s each. We calibrated the visibilities against the observation of HD 184349, a $m_K = 3.47$\,mag unresolved K4 giant star (K-band diameter of $1.09 \pm 0.05$\,mas following \citealt{1998A&A...331..619P}). The fit for the $2.10-2.29\, \mu$m wavelength range with a uniform disc visibility function yields a diameter of $3.6491 \pm 0.0007$\,mas with a reduced $\chi^2 = 9.9$, and is thus unsatisfactory. A much better fit with a reduced $\chi^2 = 2.8$ is obtained with a single power law limb darkening disc model from \cite{1997A&A...327..199H}. It gives a diameter of $3.881 \pm 0.007$\,mas and a mean limb darkening exponent of $0.45 \pm 0.01$. The squared visibility residuals are presented in the bottom panel of Fig.~\ref{fig:xiTel}. They are typically smaller than 0.005, corresponding to less than 0.25\,\% on the visibility. The limb-darkened disc model reduced the residuals by a factor two compared to the uniform disc. Even though the star is not fully resolved, the GRAVITY data are clearly accurate enough to detect limb darkening in the first lobe of the visibility function.

\subsubsection{Depth of the first null of 24\,Cap}
\label{sec:24Cap}

\begin{figure}
\vspace{0 cm}
\centering
\includegraphics[width=\hsize]{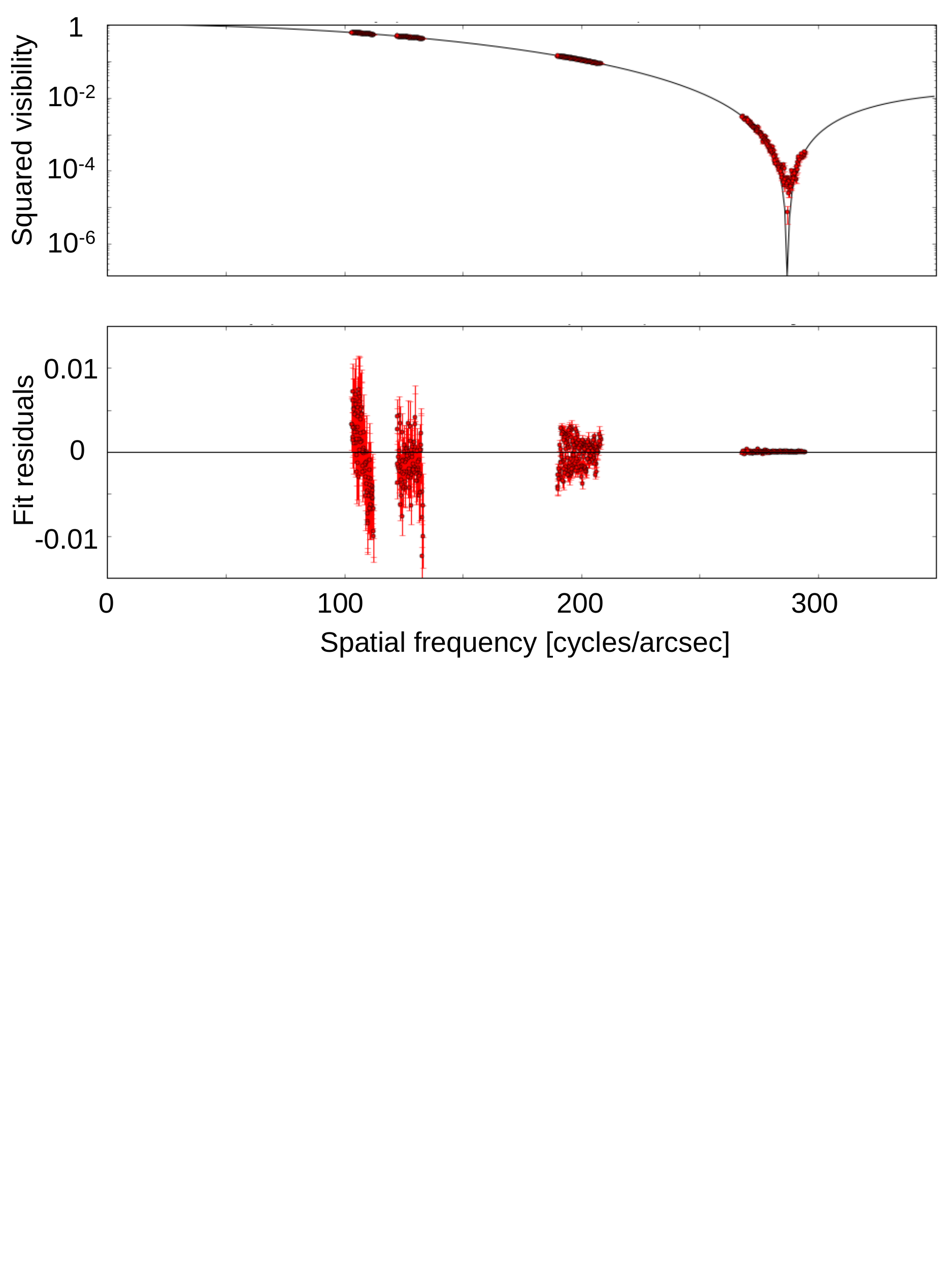}
\vspace{-6 cm}
\caption{Depth of the first null of the K5/M0 III giant 24\,Cap: the top panel shows the observed squared visibility for 24\,Cap. The depth of the null of the first minimum is $8 \times 10^{-6}$, corresponding to a visibility modulus of 0.3\,\%. The non-zero visibility is likely caused by the granulation in the star's surface. The solid line is the best fit limb darkening disc model, the fit results are shown in the lower panel.}
\label{fig:24Cap}
\end{figure}

We measured the accuracy of GRAVITY at low visibilities with the K5/M0 III giant 24\,Cap, a bright $m_K = 0.53$\,mag star (Fig.~\ref{fig:24Cap}). Each calibrated point is the combination of 30 exposures of 5\,s each. As a visibility calibrator star we used HD\,196387, a $m_K = 3.47$\,mag unresolved K4 giant star (K-band diameter estimated to $1.093 \pm 0.015$\,mas by \citealt{2005A&A...433.1155M}). Because of the high brightness of the source, the fringe tracker detector response was nonlinear by a few percent. The resulting additive errors are larger when the visibilities are high, therefore primarily affecting the low spatial frequency channels, whose visibilities get underestimated. We thus excluded the visibilities of the shortest baseline from the model fitting, and applied an ad-hoc correction by a factor 1.025 in Fig.~\ref{fig:24Cap}. We fitted the data with a single power law limb darkening disc model from \cite{1997A&A...327..199H} over the $2.10-2.29\, \mu$m wavelength range. This gives a diameter of $4.473 \pm 0.005$\,mas and a mean limb darkening exponent of $0.304 \pm 0.008$. The squared visibility residuals are presented in the lower panel of Fig.~\ref{fig:24Cap}. They are typically smaller than 0.01 (equivalent to residuals less than 0.5\,\% on the visibility at high visibility). 

The spatial frequency coverage allowed to sample the first zero of the visibility function and to measure the depth of the first null. As shown on the upper panel of Fig.~\ref{fig:24Cap}, the minimum squared visibility is $8 \times 10^{-6}$ or a visibility modulus of 0.3\,\%. This is comparable to measurements of other K giants with high accuracy instruments, e.g., the measurement of Arcturus, another K giant, with the IONIC instrument on the IOTA interferometer by \cite{2008A&A...485..561L}. The typical value of a few 0.1\,\% for the null depth is probably of astrophysical origin and explained by the granulation at the surface of the star. GRAVITY is therefore at least as good as the most accurate interferometers to remove biases on squared visibilities.

\subsection{Spectro-differential visibilities of the T\,Tauri binary S\,CrA}

\begin{figure*}
\vspace{0 cm}
\centering
\includegraphics[width=\hsize]{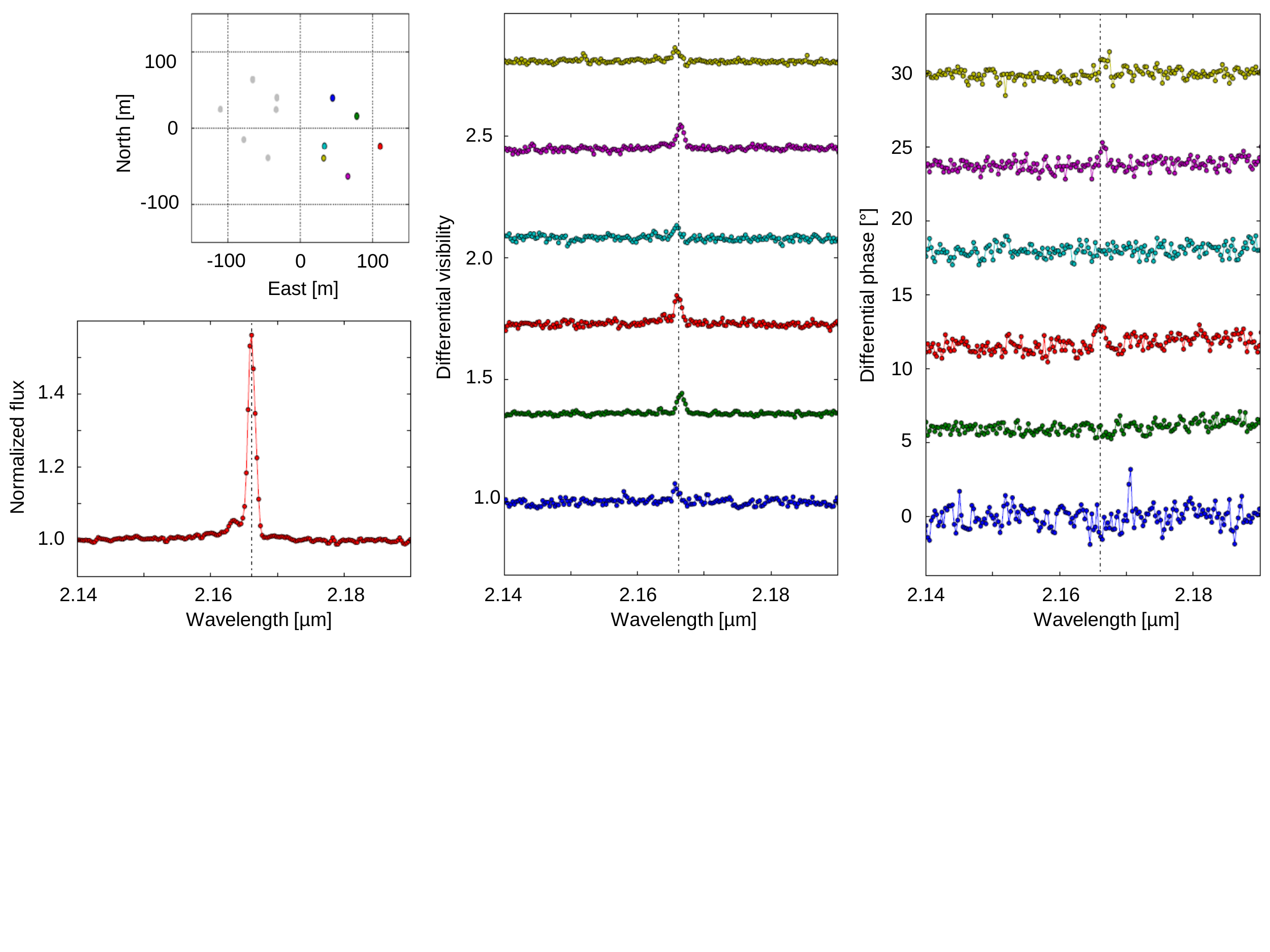}
\vspace{-4.5 cm}
\caption{Spectro-differential interferometry of the T\,Tauri star S\,CrA North: the lower left panel shows the spectrum around the double-peaked hydrogen Br$\gamma$ emission line. The middle and right panels display the observed visibilities and differential phases, respectively, for the six UT baselines depicted on the top left. The asymmetry and velocity offsets in the visibilities of the Br$\gamma$ line at short baselines and the more symmetric visibility spectrum at long baselines point towards two different structures in the ionized gas of the innermost accretion / ejection region of T\,Tauri stars.}
\label{fig:SCrA}
\end{figure*}

The physical structure and processes in the inner regions of protoplanetary discs are still poorly constrained, yet they are key for understanding planet formation. In this region, accretion flows, winds and outflows are essential to control angular momentum, alter the gas content, drive the dynamics of the gas, and set the initial conditions for forming terrestrial planets. The observations of the innermost few astronomical units require milliarcsecond resolution as provided by infrared interferometry. \cite{2015A&A...574A..41A} and \cite{2017A&A...599A..85L} have presented broad-band interferometric surveys of T\,Tauri and Herbig AeBe stars, respectively, to derive the statistical properties of the dust distribution in the inner discs. GRAVITY will be able to take the next steps and image the hot (Br$\gamma$) and warm (CO) gas in the protoplanetary discs, as well as to extend the sample to a wider range of stellar masses, ages and disc properties. 

As an illustration, we used the dual-field mode of GRAVITY to perform spectro-differential interferometry on the T\,Tauri binary system S\,CrA, which consists of an early G-type primary and an early K-type secondary star. These are the first high spectral resolution $R \approx 4500$ interferometric observations of a T\,Tauri star. S\,CrA\,North, for example, was observed with the VLTI before \citep{2012A&A...543A.162V}, but it was out of reach for spectro-interferometrically resolving its spectral lines. The binary S\,CrA is particularly well suited for GRAVITY observations with the UTs, because its separation of about 1.3\arcsec\ allows using the dual-field mode, thereby avoiding to split the light between the fringe tracker and the science spectrometer, and increasing the sensitivity by a factor two. S\,CrA North has a brightness of $m_K = 6.6$\,mag and S\,CrA South $m_K = 7.3$\,mag  \citep{2003ApJ...584..853P}. Both stars were observed at high spectral resolution on 19 July 2016 and 15 August 2016 with the four UTs. On both nights we swapped between the two components of the binary system, i.e. we first used the Southern component for fringe-tracking to observe S\,CrA North at high spectral resolution and with long integration times, and subsequently swapped the two stars to fringe-track on S\,CrA North and to perform long integrations on S\,CrA South. Each measurement consists of five exposures with 60\,s integration time. We calibrated each visibility point with on-axis observations of HD\,188787. 

Fig.~\ref{fig:SCrA} shows the observed spectrum, visibilities and differential phases of S\,CrA North around the hydrogen Br$\gamma$ line. We calibrated the absolute visibility of the continuum using the fringe tracker data recorded as part of the swapping sequence. From the Gaussian fit to the K-band visibilities, we find a half-flux radius for the continuum emission around the Br$\gamma$ line of 0.83~$\pm$~0.04\,mas, which translates to 0.108~$\pm$~0.005\,AU at a distance of 130\,pc \citep{2003ApJ...584..853P}. The Br$\gamma$ emission line of S\,CrA North is double-peaked at 2.1635\,$\mu$m and 2.1661\,$\mu$m, corresponding to radial velocities of about $-360$\,km/s and 0\,km/s, respectively. For all six baselines, the visibility increases around the peak of the line. On the other hand, the visibilities across the Br$\gamma$ emission line vary between the baselines, and a phase signal of few degrees is only detectable on three baselines. Our observations therefore suggest the presence of two structures in the environment of S\,CrA North with different kinematic signatures at small and large scales. Thanks to spectro-differential interferometry, GRAVITY thus provides new insights into the innermost regions of T\,Tauri stars, probing the complex accretion-ejection phenomena at play. 

\subsection{Imaging the core of $\eta$\,Car}

\begin{figure*}
\vspace{0 cm}
\centering
\includegraphics[width=21 cm]{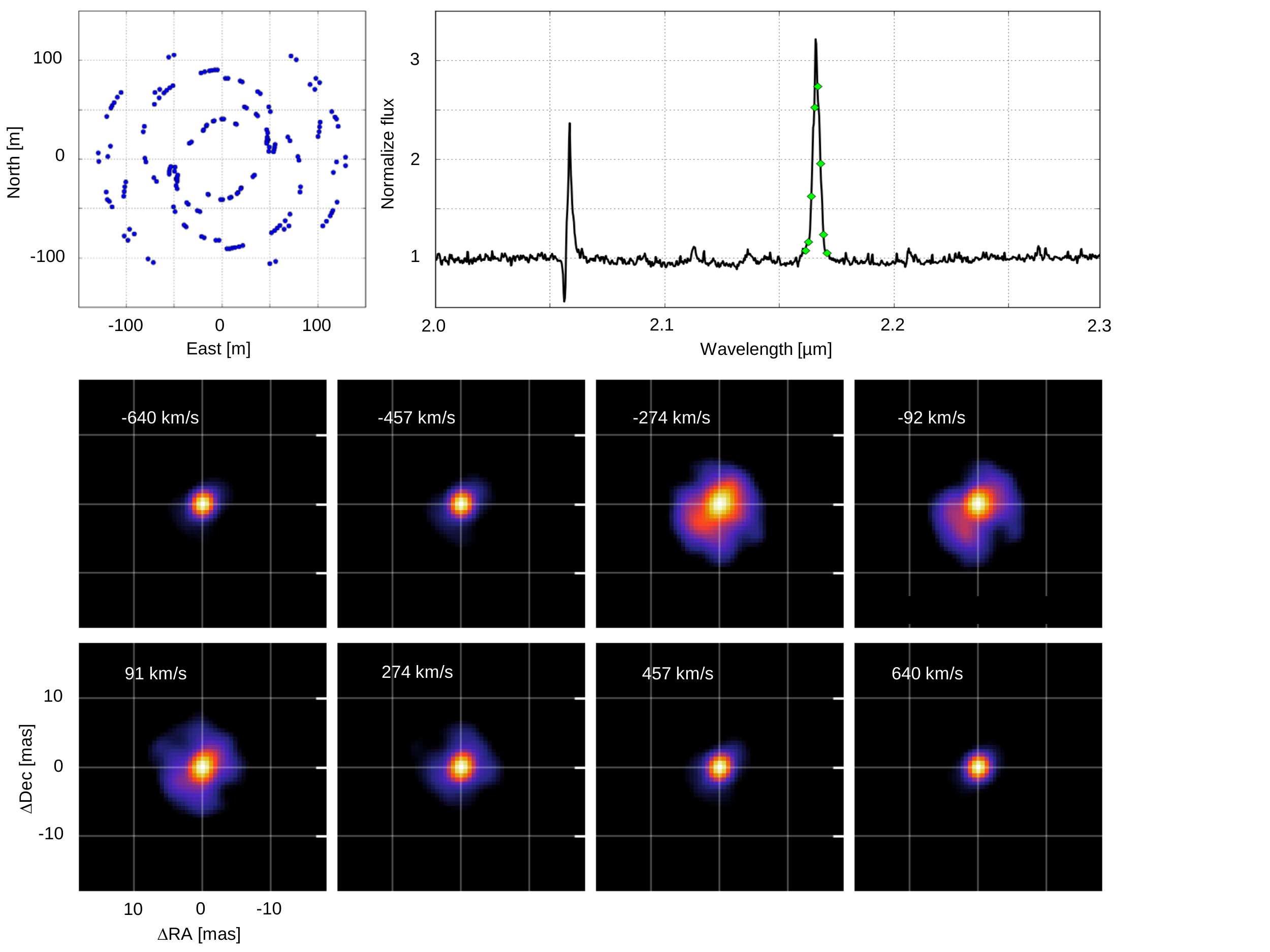}
\vspace{0 cm}
\caption{Core of $\eta$\,Car at mas resolution: The top left panel displays the baseline coverage of our observations with the ATs. The top right panel shows the observed spectrum around the \ion{He}{i} and Br$\gamma$ emission lines. The lower panels present the reconstructed images for different wavelengths (indicated by the green diamonds in the spectrum) across the Br$\gamma$ line. The frames are individually normalized, the minimum brightness corresponds always to 5\,\% of the maximum brightness. The images reveal the complex morphology of the primary wind and its interaction with the wind from the hidden secondary star.}
 \label{fig:EtaCar}
\end{figure*}

The relative proximity of $\eta$\,Car makes it one of the best observable high-mass stars in the Galaxy. GRAVITY enables spectrally resolved interferometric imaging of the central wind region of this compact binary. At an apparent brightness of $m_K = 0.94$\,mag, this target is bright enough for observations with the ATs, which can be relocated to different configurations and thereby provide the rich baseline coverage needed for high fidelity interferometric imaging. We observed $\eta$\,Car at high-spectral resolution ($R \approx 4500$) and split-polarization on 25 and 28 February 2016, and obtained 10 data sets with the A0-G1-J2-K0 telescope configuration, plus four more data sets with the A0-G2-J2-J3 configuration. Each data set consists of 30 exposures with 10\,s integration time. The top left panel of Fig.~\ref{fig:EtaCar} displays the baseline coverage of the observations. With a maximum baseline $B$ of $\sim 130$\,m and minimum baselines of $\sim 40$\,m, we obtain angular resolutions $\theta=\lambda/2B \approx 1.7-5.6$\,mas at the wavelength of Br$\gamma$, respectively. The squared visibilities have an average signal-to-noise ratio of 23 and the closure phases exhibit an average error of 5.5$^\circ$. The calibrated spectrum (top right panel of Fig.~\ref{fig:EtaCar}) contains several lines, the most prominent corresponding to \ion{He}{i} 2s-2p (2.059\,$\mu$m) and Br$\gamma$ (2.166\,$\mu$m). 

The high accuracy visibilities (see \ref{HighAccuracyVisibilityObservationsOfResolvedStars}) and the high signal-to-noise ratio allow a chromatic image reconstruction of $\eta$ Car across the Br$\gamma$ line. For this we used SQUEEZE \citep{2010SPIE.7734E..2IB}, an interferometry imaging software that allows fitting simultaneously the squared visibilities, closure phases, and chromatic differential phases. Compared to radio interferometric observations, the few telescopes combined in optical interferometry provide only a sparse baseline coverage, and the image reconstruction is usually guided with so-called regularizers. In our case we used a combination of the L0-norm to avoid spurious point-like sources and a Laplacian to enhance the extended structure expected in some channels. In addition we included a spectral regularizer that computes the L2-norm across the different spectral channels to ensure spectral continuity across the emission line. SQUEEZE uses a Simulated Annealing Monte-Carlo algorithm as the engine for the reconstruction. In our case, we created 30 chains with 250 iterations each to find the most probable image. We used a 167$\times$167\,pixel grid with a scale of 0.6\,mas/pixel, and chose a Gaussian with 50\,\% of the total flux centered in the image as the starting point for the reconstruction. The final images are the average of the frames from the different Monte-Carlo chains that converged to a reduced $\chi^2 \sim 1$. The bottom panels of Fig.~\ref{fig:EtaCar} display eight images across the Br$\gamma$ line, each convolved with a Gaussian with a full-width-half-maximum of 1.76\,mas. 

The reconstructed images reveal the different wind components of $\eta$ Car. The first and last frames, close to the continuum level, show a compact structure with a size of $\sim 5$\,mas. This compact component is the optically-thick primary wind reported previously with interferometric observations \citep{2003A&A...410L..37V, 2006SPIE.6268E..2SW}. It is slightly elongated with a position angle of $\sim 134^\circ$, which is coincident with the orientation of the major-axis of the homunculus (position angle $\sim 131^\circ$). The cavity created by the wind-wind collision between the two binary components \citep{2013MNRAS.436.3820M} is seen at a velocity of $\sim -275$\,km/s. At this velocity, $\eta$\,Car shows an asymmetric elongated structure with a size of $\sim 15$\,mas, with a major axis in the South-East direction. This result is consistent with the fan-shaped morphology found by \cite{2016A&A...594A.106W} for similar velocities in the AMBER/VLTI data. The morphology of the Br$\gamma$ blue wing could be also affected by an additional contribution from the fossil winds at the core of $\eta$\,Car \citep{2011ApJ...743L...3G, 2016MNRAS.462.3196G}. In contrast, the red wind exhibits a more compact structure, because the emitting region corresponds mainly to the back side of the primary wind. As illustrated by this example, the spectro-interferometric capabilities of GRAVITY combined with its high-sensitivity and performance provide unique tools for chromatic image reconstruction to reveal the milliarcsecond morphology of a large variety of astrophysical targets.

\subsection{Microarcsec spectro-differential astrometry}

In the examples of resolved stars (Section \ref{HighAccuracyVisibilityObservationsOfResolvedStars}), we have concentrated on the absolute accuracy of the visibility modulus. When it comes to position measurements, however, the information is better traced by the interferometric phase. For example, a phase difference of 1$^\circ$ at a wavelength of 2.2\,$\mu$m and a baseline of 130\,m corresponds to an angular displacement of 10\,$\mu$as. This kind of accuracy is easily achieved by GRAVITY when looking at the phase difference over a narrow wavelength range. In the following we demonstrate such 10\,$\mu$as spectro-differential astrometry on the examples of tracing the wind or gas stream in the high-mass X-ray binary (HMXB) BP\,Cru and the broad line region of the quasar PDS\,456.

\subsubsection{Tracing the inner region of the HMXB BP\,Cru}

\begin{figure*}
\vspace{0 cm}
\centering
\includegraphics[width=\hsize]{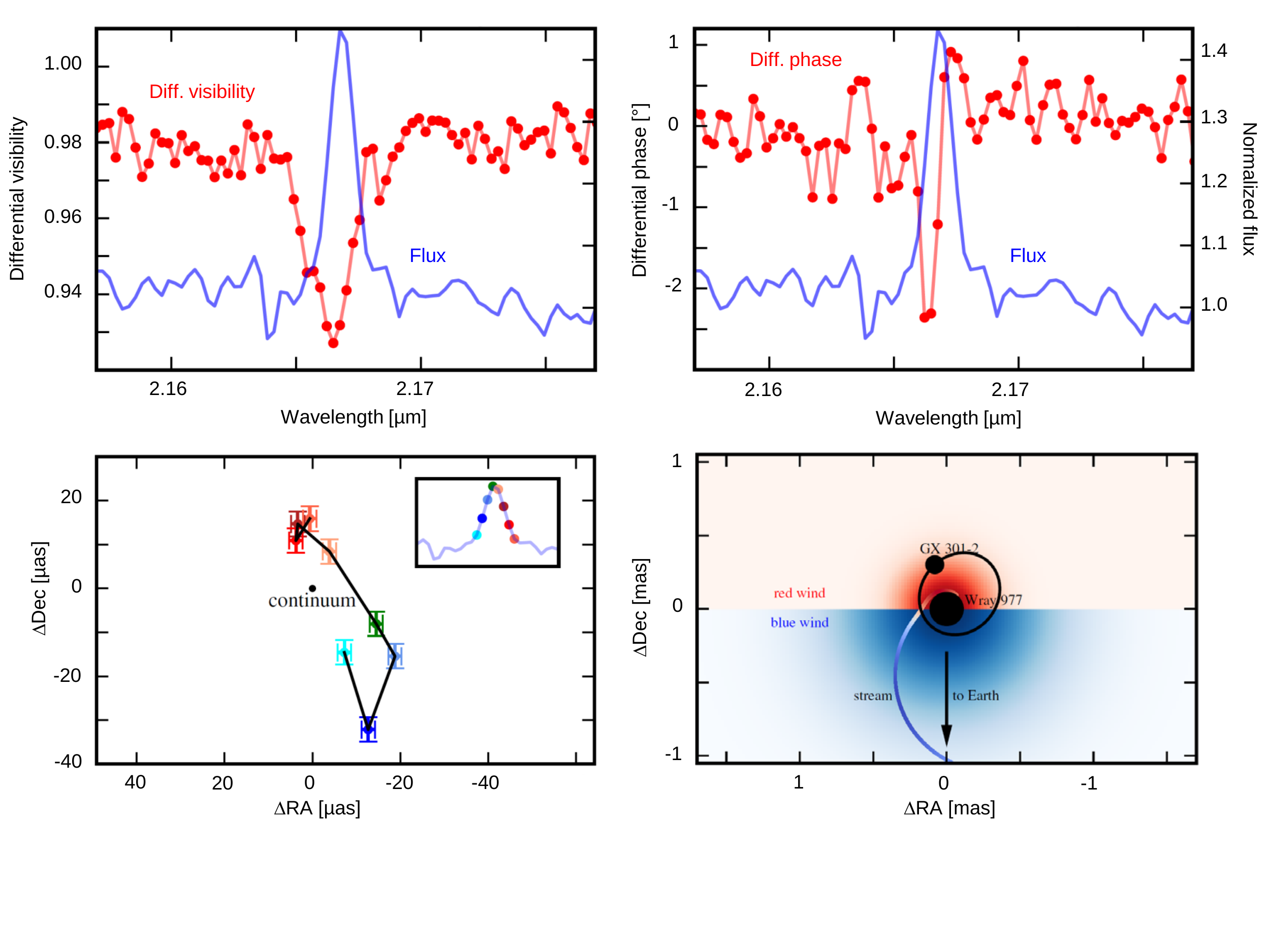} 
\vspace{-2 cm}
\caption{Tracing the inner region of the HMXB BP\,Cru: the top figures display the differential visibility amplitudes (left) and phases (right) across the Br$\gamma$ line (red) and the continuum normalized spectrum (blue) exemplary for one of the six observed baselines. The lower left panel shows the model-independent centroid positions for each wavelength across the Br$\gamma$ line. The image on the blue side of the line has a larger centroid shift than the image on the red side. Note the $\mu$as scale on the lower left panel. The schematic on the lower right panel shows the BP Cru system as inferred from simple geometric model fits to the interferometric signatures. The donor star, the X-ray pulsar and its binary orbit are shown in black. An extended wind could explain the visibility amplitude drops across the emission line. A size asymmetry between the blue and red shifted sides of the wind could be related to the X-ray illumination by the compact object. Non-zero visibility phases, also asymmetric between blue and red wavelengths, could be explained by the presence of a gas stream. A simple stream model following \cite{2008MNRAS.384..747L} is shown in the bottom right, with the colors along the stream corresponding to the radial velocity measured at Earth ($i=60^\circ$).}
\label{fig:BPCru}
\end{figure*}

With typical orbital size scales $< 1$\,mas, X-ray binaries are beyond the imaging resolution of optical/near-infrared interferometers. As a result, spatial information about the accretion process or binary interaction is typically inferred from photometry and spectroscopy. For systems containing spectral lines, however, spectro-differential interferometry may be used to achieve differential astrometry on the few\,$\mu$as scale between continuum and line emission. This technique builds on the exquisite differential visibility and phase precision, which GRAVITY achieves by virtue of its fringe tracker and the possibility for minute long coherent integrations at high spectral resolution. 

Here we illustrate this technique with the observations of the high-mass X-ray binary BP\,Cru using the four UTs on the night of 18 May 2016. The total on-source integration time was 2100\,s with individual exposures of 30\,s. We took all data in high spectral resolution ($R \approx 4500$). BP\,Cru is a canonical HMXB, in which the massive ($> 1.85\,M_\odot$), slowly rotating ($P = 696$\,s) pulsar GX\,301-2 accretes from the strong stellar wind of the early-blue hypergiant Wray\,977 \citep{2006A&A...457..595K} along an eccentric orbit ($e=0.462$). X-ray light curves and column densities, however, show evidence for a more complex accretion mechanism, which likely includes a gas stream of enhanced density \citep{1991ApJ...376..245H, 2008MNRAS.384..747L}. Furthermore, X-ray photoionization and heating could substantially affect the side of the stellar wind facing the pulsar \citep{1994ApJ...435..756B, 2015A&A...575A...5C}. Therefore spatially resolving the inner region of this system could provide important information about the gravitational and radiation effects of the compact object on the stellar companion. 

The K-band spectrum of BP\,Cru obtained with GRAVITY shows prominent emission lines of \ion{He}{i} ($2.059\,\mu$m) and Br$\gamma$ ($2.166\,\mu$m), both of which show differential visibility amplitudes and phase signatures across the line. Fig.~\ref{fig:BPCru} shows exemplary the Br$\gamma$ line for the UT1-UT4 baseline. We note the typical standard deviation in the continuum is $\sim 0.4$\,\% and $\sim 0.2^\circ$ for amplitudes and phases, respectively, the latter of which corresponds to 2\,$\mu$as for a $100$\,m baseline. These signatures can be used in the framework of the marginally resolved limit \citep{2003A&A...400..795L} to derive model-independent information about the moments of the flux distribution. They point to velocity dependent centroids and extension across the line, which cannot be explained by a symmetrical, undisturbed stellar wind. Fig.\ref{fig:BPCru} (bottom) shows the centroids of the image in the sky plane as a function of wavelength across the line. The blue side must be more displaced from the continuum than the red side, and they are in opposite directions. 

The fitting of simple geometrical models suggests an extended, distorted wind observed on scales a few times the size of the orbit. Alternatively the differential phases could be explained by a compact, linear structure with orbital size scale, potentially associated with a gas stream from the donor star. Future spectroscopic observations at high resolution, coupled with interferometric observations at different orbital phases, will help to distinguish between more complex models. 

\subsubsection{Broad line region of the quasar PDS\,456}

\begin{figure}
\vspace{0 cm}
\centering
\includegraphics[width=\hsize]{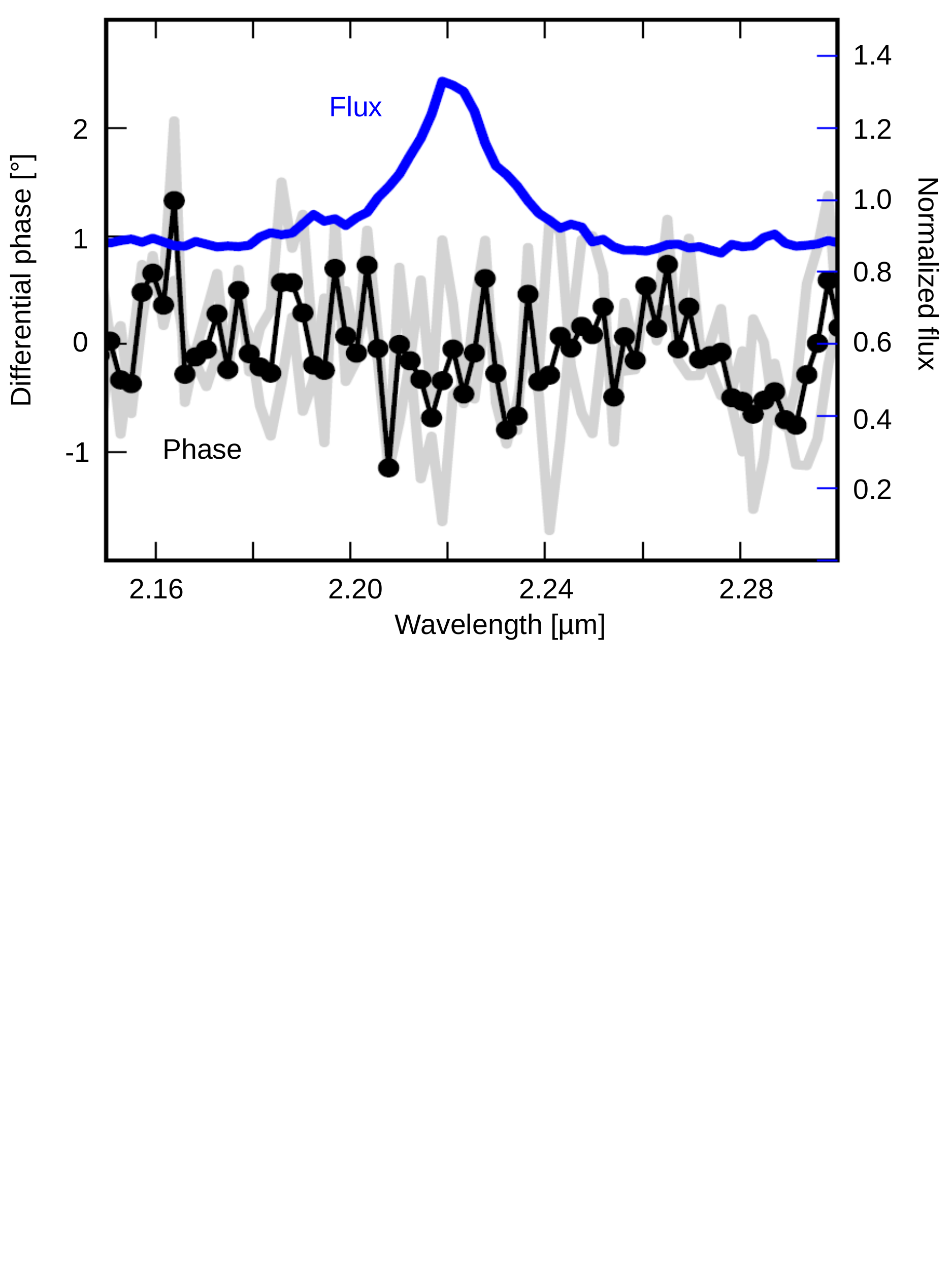}
\vspace{-6 cm}
\caption{Broad line region of the quasar PDS\,456: the plot shows the normalized spectrum (blue) and differential phase (black) of the broad Pa$\alpha$ line for the average of the UT1-UT4 and UT2-UT4 baselines (individual baselines in light gray). No differential phase signature is detected, neither in the individual baselines nor in their average, despite a precision of $\lesssim 1^\circ$.}
\label{fig:PDS456}
\end{figure}

What is the degree of ordered vs. virial motion of the broad line region (BLR) in active galactic nulcei (AGN)? The answer is key to improving the mass-luminosity and size-luminosity relations of AGN, which in turn are cornerstones for using quasars as cosmological standard beacons and to trace the co-evolution of massive black holes and their host galaxies. If the degree of ordered rotation is high, this rotation should be reflected in a velocity dependent displacement of the order of a few 10 $\mu$as for the brightest AGN \citep{2015MNRAS.447.2420R}. 

The astrometric signature is expected to be particularly large for the strong Pa$\alpha$ line, but this line is only observable in K-band for a redshift larger than $z \gtrsim 0.07$. We thus selected the quasar PDS\,456 ($m_K = 9.9$\,mag, $m_V = 14.0$\,mag), a $z = 0.184$ near-Eddington X-ray source \citep{2000MNRAS.312L..17R}, to observe with GRAVITY and the UTs using the visible adaptive optics MACAO on 19 July 2016. We used the medium spectral resolution ($R \approx 500$) with exposure times of 30\,s per frame, and spent a total of 3900\,s on source. We detected significant coherent flux on all baselines, despite problems with the fringe-tracking because of the limited MACAO performance at faint $V$ magnitudes. The continuum appears partially resolved, with a best Gaussian fit FWHM size of $\sim 0.3$\,mas, comparable to the angular size expected from the relation between dust sublimation radius and AGN luminosity \citep{2015ARA&A..53..365N}. This size is much smaller than the VLTI beam and subject to systematic errors in calibration, and so we cautiously interpret it as an upper limit of 0.6\,mas. 

Fig.~\ref{fig:PDS456} shows the observed spectrum and differential phase. The broad Pa$\alpha$ line -- redshifted to the K-band -- has a full width half maximum (FWHM) of 3200 \,km/s. No differential phase or amplitude signature is detected at the Pa$\alpha$ line, despite a precision of $\lesssim 1^\circ$ on each of the two longest baselines. This constrains the line centroid displacement to $\lesssim 10\,\mu$as. Using the measured line strength and assuming ordered rotation or outflow, we place an upper limit on the offset of the line emission from the continuum of $\lesssim 150\,\mu$as. Assuming the broad line is produced at the same spatial scale as the continuum emission, $\lesssim 50$\,\% of the line flux could be produced by a component with ordered velocities. Given the phase noise demonstrated by these observations, future GRAVITY observations should be able to probe the structure of the broad line regions of many AGN (e.g., \citealt{2015MNRAS.447.2420R}).

\subsection{GRAVITY observations of the Galactic Center}
\label{GRAVITYObservationsOfTheGalacticCenter}

The main scientific driver for GRAVITY is the supermassive black hole in the Galactic Center (e.g., \citealt{2010RvMP...82.3121G} and references therein), and addresses several fundamental questions in physics and astrophysics. These include: does Einstein's theory of general relativity correctly describe the dynamics and light propagation in its vicinity? What is the exact mass and what is the spin of the black hole? Do the stars follow simple general-relativistic orbits around a single central mass, or are the effects from encounters with dark objects or from an extended mass distribution stronger than the general relativistic effects? How do black holes accrete, and which physics is at work in the inner accretion and outflow region? Is there a jet ejected from the black hole? The GRAVITY collaboration will spend the largest fraction of its guaranteed observing time with the UTs to tackle these questions. 

\begin{figure*}
\vspace{0 cm}
\centering
\includegraphics[width=\hsize]{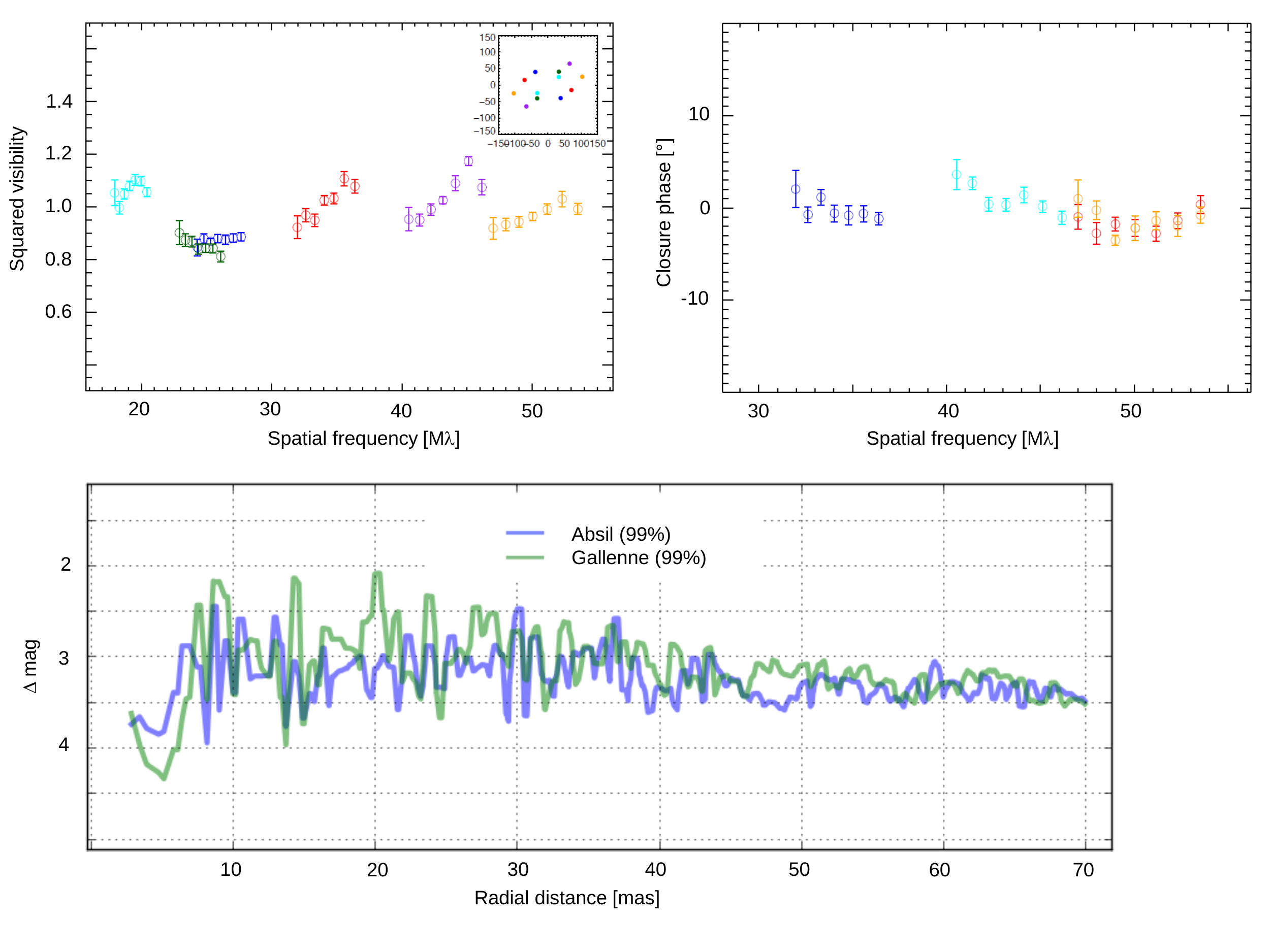} 
\vspace{-0.5 cm}
\caption{Galactic Center star S2: The top panel shows the observed squared visibilities and closure phases. The baseline coverage is shown in the small inset. The bottom panel displays the $3\,\sigma$ detection limit for companions of S2 as function of separation applying two different methods \citep{2011A&A...535A..68A,2015A&A...579A..68G}. Companions brighter than $\Delta m_K =3$\,mag, corresponding to a limiting magnitude of $m_K = 17.1$\,mag, are ruled out.}
\label{fig:S2}
\end{figure*}

\begin{figure*}
\vspace{0 cm}
\centering
\includegraphics[width=17.8cm]{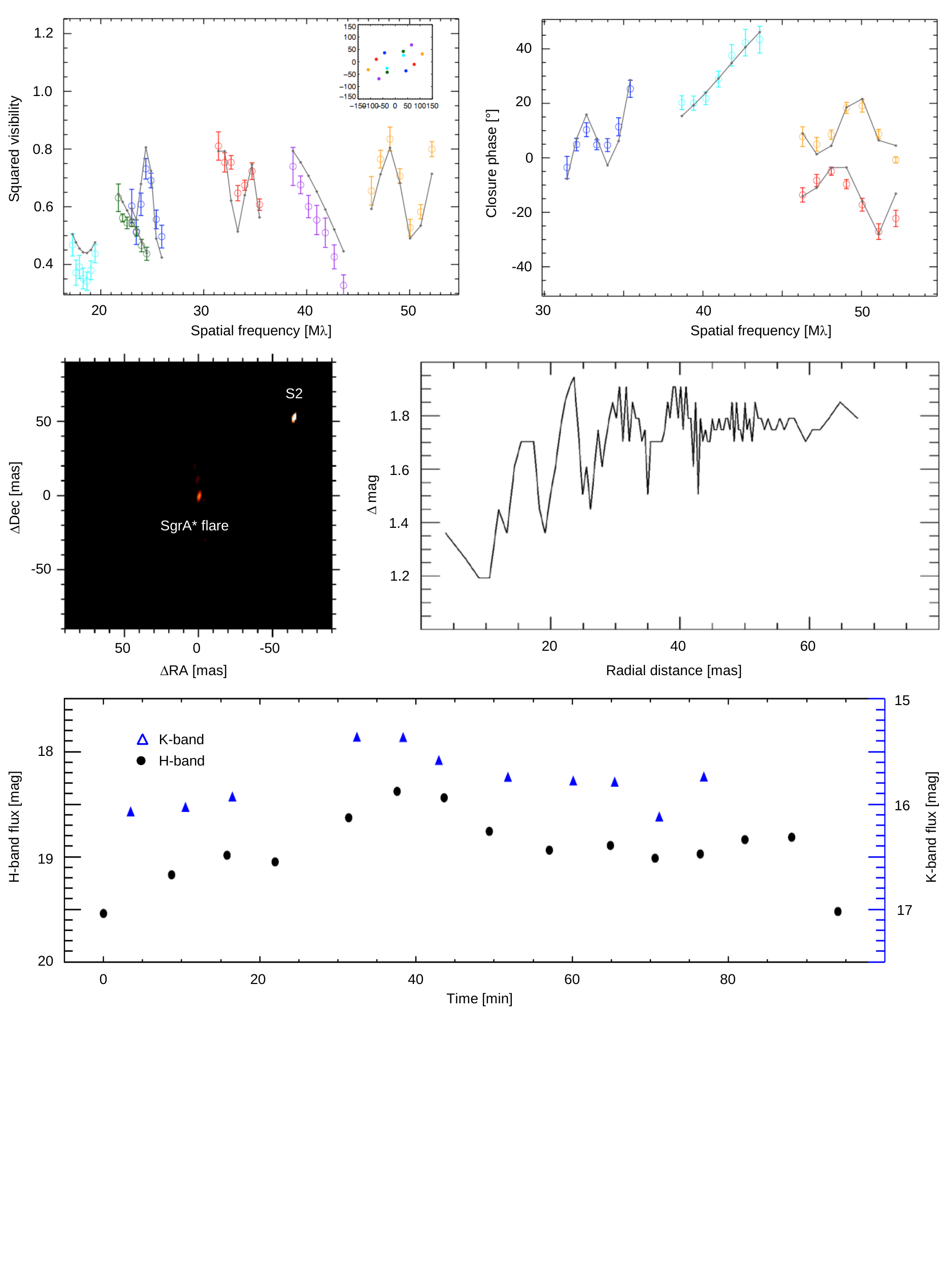} 
\vspace{-4.5 cm}
\caption{First detection of Sgr\,A* in infrared interferometry: the top panels show the observed squared visibilities and closure phases for a 5~minute exposure centered on Sgr\,A* during a flare with a peak brightness of about $m_K \approx 15$\,mag. The prominent modulation in the two quantities results from the two brightest objects in the field, the flaring Sgr\,A* and the star S2. The black line is the best fit ``binary'' model. The middle left panel displays the reconstructed image from the combination of three exposures around the peak of the flare. The middle right panel plots the 3\,$\sigma$ detection limits for a third source in the full data set, for which the average flare brightness was $m_K \approx 15.5$\,mag. We can exclude a third source brighter than $m_K = 17.1$\,mag for 90\,\% of the area. The lower panel shows the H-band (black) and K-band (blue) lightcurve of Sgr\,A* during the flare, as measured from the acquisition camera images and derived from the best fit to the interferometric visibilities and closure phases, respectively.}
\label{fig:SgrA}
\end{figure*}

\subsubsection{Observing strategy and first observations}

GRAVITY employs two main observing modes on the Galactic Center. First, interferometric imaging of the central region to (a) precisely measure the orbital parameters of the star S2 relative to the flaring black hole, during the years when S2 is closest to Sgr\,A* and both are within the same interferometric field of view, to (b) trace the motion of the flare on short time scales, which is expected if the emission originates from a small region on a close orbit or in a jet, and to (c) search for fainter stars in the central light week and follow their short period orbits. The second observing mode is long-term astrometric monitoring of the S-stars and the flaring black hole relative to more distant reference stars on very long period orbits. These data would also characterize the binary nature and winds of the reference stars, which are members of the population of young massive stars in the Galactic Center. Both types of observation can be executed simultaneously thanks to the dual-beam design of GRAVITY.

Here we report on our first Galactic Center observations with the full GRAVITY instrument on 20 and 21 September 2016, shortly after the installation of all four CIAO systems\footnote{We obtained first fringes of S2 already on 18 May 2016. See \url{https://www.eso.org/public/news/eso1622/}. At that time our CIAO infrared wavefront sensors were not yet fully installed, and we took advantage of exceptionally good seeing conditions, using the visible light wavefront sensor MACAO.}. The atmospheric conditions were slightly better than the median with an optical seeing $\sim 0.7$\arcsec\ and a coherence time of 4 to 5\,ms. We used the bright supergiant GC\,IRS\,7 ($m_K \approx 6.5$\,mag), located 5.5\arcsec\ north of Sgr\,A* for the wavefront sensing with CIAO. The observing sequence was as follows: first we used the brightest WR/O star in the field, GC\,IRS\,16NE ($m_K = 8.9$\,mag, \citealt{2006ApJ...643.1011P}), for fringe-tracking and observed GC\,IRS\,16C ($m_K = 9.7$\,mag) with the science beam combiner in low resolution mode ($R \approx 22$). Then we swapped the two stars, i.e. we used GC\,IRS\,16C for fringe-tracking and observed GC\,IRS\,16NE with the science beam combiner. This swapping provides an efficient cross-calibration of the visibilities and at the same time gives the astrometric zero point for the laser metrology. After the swapping sequence we fed S2 to the science beam combiner, while GC\,IRS\,16C remained the fringe-tracking star. On 21 September 2016 we extended the observing sequence with a blind offset to the position of Sgr\,A*. We noticed a moderately bright flare from Sgr\,A* on the acquisition camera and decided to stay on the object until we reached the compensation limit of the VLTI delay lines. After that we observed the sky frames.

 \subsubsection{No stars brighter than $m_K = 17.1$\,mag near S2}
 \label{sec:S2}

The star S2 is one of the stars with the closest known peri-center distance to Sgr\,A*. It orbits Sgr\,A* with a period of 16 years \citep{2002Natur.419..694S, 2008ApJ...689.1044G, 2017ApJ...837...30G} and reaches a velocity of up to 7600\,km/s, i.e. 2.5\% the speed of light, which makes it the prime candidate to study general relativistic orbit effects close to a supermassive black hole. At an apparent magnitude of $m_K=14.1$\,mag, S2 was far too faint for interferometry with all previous instruments. The observations of such faint objects are now possible with GRAVITY, because its fringe tracker can also work off-axis, and the stable performance allows increasing the coherent integration time by orders of magnitude. 

We observed S2 with 10\,s exposures at low spectral resolution ($R \approx 22$), the total integration time on source was 300\,s. We debiased the data with a matching 300\,s sky observation, and calibrated the visibilities with GC\,IRS\,16C observed with the same settings. The resulting visibilities and closure phases\footnote{Throughout the paper we use the longest baseline of the triangle to denote the spatial frequency of the closure phases.} of S2 are featureless (see Fig.~\ref{fig:S2}). The closure phase scatters around zero with $1.7^{\circ}$ rms. This is only marginally higher than the average closure phase error of $1^{\circ}$ obtained from boot-strapping. This strongly indicates an unresolved source. The squared visibility shows more substructure than the closure phase. This is not surprising, because the source is quite faint and therefore very sensitive to additive noise or imperfect bias subtraction, and the visibility is close to one and therefore very sensitive to the multiplicative noise from atmospheric piston residuals and coupling fluctuations. The mean squared visibility is $0.97 \pm 0.09$. There is no apparent drop at large spatial frequencies, which is again consistent with an unresolved source. 

Because of the tidal field from Sgr\,A*, S2 cannot have physical companions that could be resolved with GRAVITY. The maximum separation for such companions is set by the tidal disruption radius, which for S2 is 0.24\,mas, i.e. smaller than the resolution of GRAVITY. The extremely dense stellar environment, however, can lead to chance associations with other stars in the field. We use two methods to determine the detection limits for such faint companions: the first method introduced by \cite{2011A&A...535A..68A} assumes that the data follows Gaussian statistics and compares the $\chi^2$ of the null hypothesis with a companion model. The second method -- CANDID \citep{2015A&A...579A..68G} -- injects fake companions in the data and tries to recover them. Fig.~\ref{fig:S2} shows the resulting limits as function of separation. We can rule out companions brighter than $\Delta m_K =3$\,mag, i.e $m_K < 17.1$\,mag at a significance level of $3\,\sigma$. The limiting magnitude of $m_K = 17.1$\,mag corresponds to a B8V star at the Galactic Center distance of 8.3\,kpc \citep{2017ApJ...837...30G} and K-band extinction of 2.7\,mag \citep{2011ApJ...737...73F}. Overall our observations already lead to a firm upper limit despite the short observing time and the consequently limited baseline coverage. Longer observations with a better baseline coverage will quickly lead to tighter constraints.
 
\subsubsection{First detection of Sgr\,A* in infrared interferometry}

The compact radio source Sgr\,A* marks the position of the Galactic Center black hole. The radio emission originates from the hot, ionized plasma of the inner accretion zone. On top of this quasi-steady component there is variable emission in the X-ray and infrared bands, which appears as flares, typically a few times per day and lasting for about $1-2$\,h \citep{2001Natur.413...45B,2003Natur.425..934G, 2004A&A...427....1E,2005ApJ...628..246E}. The flares originate from transiently heated electrons of the inner accretion / outflow region \citep{2009ApJ...698..676D}, but the details are still under debate. Because any matter close to the black hole moves with a sizable fraction of the speed of light, GRAVITY will observe a motion of the flare, if a single small region dominates its emission. If originating from close to the last stable orbit, the time variability and motion of the flares are directly tied to the mass and spin of the black hole. But it could also be that the flares come from a compact jet, thereby tracing the outflow dynamics of a supermassive black hole. If the flares reflect statistical fluctuations in the accretion flow, GRAVITY will observe no or random motion \citep{2014MNRAS.441.3477V}. In any case the flares from the Galactic Center are key to the understanding of the black hole's accretion physics, and thus a prime target for GRAVITY.

Here we report on the first interferometric detection of an infrared flare from Sgr\,A* on 21 September 2016. We followed the flare for a bit more than one hour until reaching the limit of the VLTI delay lines. The peak apparent brightness during our interferometric observations was $m_K \approx 15$\,mag, the average brightness was about 0.5\,mag fainter. The fringes from the flare are clearly detected in individual 10\,s exposures. In total we could acquire 11 good data sets in low spectral resolution ($R \approx 22$), each containing 30 times 10\,s exposures. We spent a total of 3300\,s on source. The top panels of Fig.~\ref{fig:SgrA} show the visibilities and closure phases for one of the data sets. The prominent modulation of the visibility and closure phase with baseline length is the interferometric signature of the two brightest objects in the field, the flare from Sgr\,A* and the star S2. The black line shows the best fit model assuming two point sources. We repeat this ``binary'' fit for each of the 11 interferometric data sets to derive the K-band brightness ratio between Sgr\,A* and S2. This interferometric K-band lightcurve (Fig.~\ref{fig:SgrA} bottom) of Sgr\,A* follows closely the H-band light curve as measured from the acquisition and guiding camera images. 

In order to uncover a potential substructure or additional point sources, we applied several image reconstruction algorithms to create maps of the surrounding of Sgr\,A*, including CLEAN \citep{1974A&AS...15..417H}, MiRA \citep{2008SPIE.7013E..1IT} and SQUEEZE \citep{2010SPIE.7734E..2IB}. The various algorithms consistently recover Sgr\,A* and S2 in the images. Both objects are unresolved at the about 2\,mas~$\times$~4\,mas angular resolution of the VLTI. The middle left panel of Fig.~\ref{fig:SgrA} gives the example of the image reconstructed with MiRA for the combination of the 3 exposures of 5 minutes each around the peak of the flare. The spurious sources around Sgr\,A* and S2 are likely artefacts from the comparably sparse baseline coverage. 

To quantify the detection limit for a third point source, we derive for each image point the necessary brightness for a 3\,$\sigma$ detection of an artificial star when fitting a three object model to the full data set. The radial profile of the required contrast with Sgr\,A* is shown in the middle right panel of Fig.~\ref{fig:SgrA}. Taking into account the mean flare brightness of $m_K \approx 15.5$\,mag during our observations, we can exclude a third source brighter than $m_K \approx 17.1$\,mag for 90\,\% of the area within 50\,mas radius around Sgr\,A* at the time of our observation. This limiting magnitude is quite comparable to the 3 $\sigma$ limit for the non-detection of chance companions of S2 in Section \ref{sec:S2}. 
 
\subsection{Dual-beam astrometry of the M-dwarf binary GJ\,65}

One of the key scientific applications of GRAVITY is narrow-angle astrometry. This technique is based on the simultaneous observation of two objects within the VLTI field-of-view (2\arcsec\ and 4\arcsec\ with UTs and ATs, respectively). The differential nature of the measurement removes most of the astrometric uncertainties. In principle this technique is only limited by the residual atmospheric disturbance between the two objects, which averages out with observing time. Among the first to recognize the potential of narrow-angle astrometry was \cite{1980A&A....89...41L}. The application in interferometry was later proposed by \cite{1992A&A...262..353S}. They especially noted the possibility of micro-arcsecond astrometry using long baseline interferometry. In order to test and develop the GRAVITY dual object astrometry, we observed a number of binaries with separations of a few arcseconds and magnitudes $m_K \approx 1 - 7$\,mag. 

The well studied M-dwarf binary GJ\,65\,AB (BL Cet+UV Cet,  WDS J01388-1758AB) consists of two very low mass M5.5Ve and M6Ve main-sequence dwarfs with masses of $M(A)=0.123\,M_\odot$ and $M(B)=0.120\,M_\odot$ \citep{2016A&A...593A.127K}. The two stars orbit each other with a period of $P=26.28\,$yrs and a semi-major axis  $a=2.05\,\arcsec$. With a distance of only 2.68\,pc GJ\,65 is one of the closest stars to the sun. The close proximity allowed \cite{2016A&A...593A.127K} to measure the size with the VLTI. The best fit uniform disc diameter are $\theta(A) =0.56$\,mas and $\theta(B) =0.54\, \rm mas$. Both components have almost equal brightness with $m_K (A)=6.02$\,mag and $m_K (B)=6.17\,mag$. They belong to the family of flare stars, which show violent outbursts of X-ray and UV radiation. Flares can significantly increase the UV brightness of M5V-stars, yet the K-band contribution is negligible at $< 1\%$ \citep{2012ApJ...748...58D}. 

\begin{figure}
\vspace{0 cm}
\centering
\includegraphics[width=\hsize]{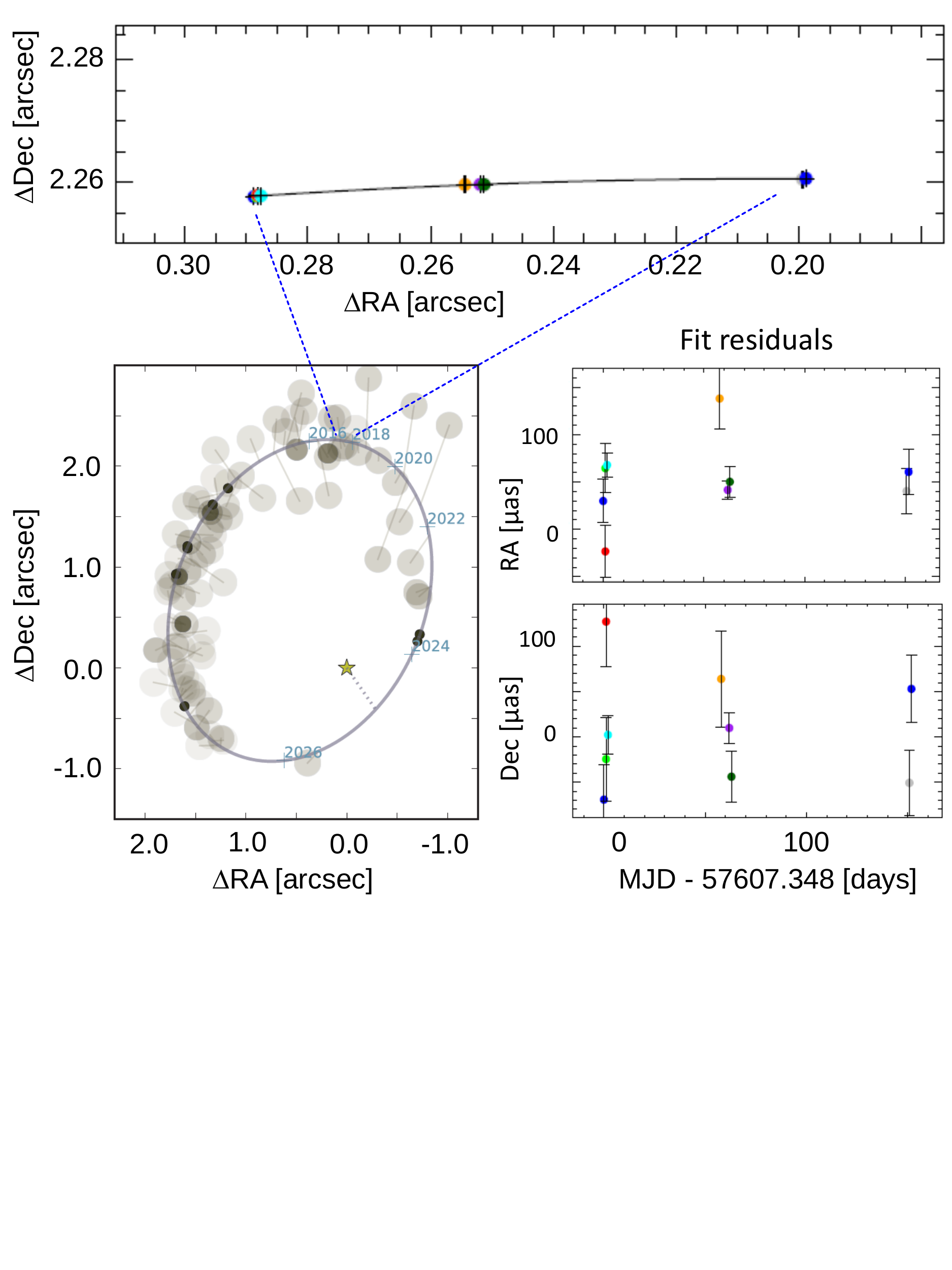} 
\vspace{-3.5 cm}
\caption{Dual-beam astrometry of the M-dwarf binary GJ\,65: The top panel shows the astrometry from three observing epochs in 2016 and the best quadratic fit (black) to the GRAVITY data. The lower left panel displays the orbit of GJ\,65 from \cite{2016A&A...593A.127K} based on seeing limited and adaptive optics observations. On the lower right we plot the residuals for the best quadratic fit to the GRAVITY astrometry. The mean rms scatter is 42\,$\mu$as and 64\,$\mu$as in right ascension and declination, respectively.} 
\label{fig:GJ65}
\end{figure}

We observed GJ\,65\,AB repeatedly with the ATs in the A0-G1-J2-K0 configuration in August 2016, October 2016 and January 2017. Fig.~\ref{fig:GJ65} shows the measured relative positions. We clearly detect an orbital motion, which we approximate with a quadratic fit to the positions. We derive the following motion: $\Delta \alpha (t) = 287.140\pm0.013\,\text{mas} - (597.54\pm0.45\,\mu\text{as/d}) \times (t - t_0) + (0.026\pm0.005\,\mu\text{as/d}^2) \times (t - t_0)^2$ and $\Delta \delta (t) = 2258.142\pm0.028\,\text{mas} + (35.92\pm0.97\,\mu\text{as/d}) \times (t - t_0) - (0.132\pm0.006\,\mu\text{as/d}^2) \times (t -t_0)^2$, with $t_0 = 57607.348$ (MJD). The individual fit errors are between $20-30\,\mu$as, about a factor 2-3 smaller than the actual scatter.

In principle, some of the scatter could be due to faint companions. However the day-to-day scatter is about as large as the month-to-month scatter, therefore the residual scatter originates more likely from multiple systematic error sources. A detailed error analysis of the GRAVITY narrow-angle astrometry has been presented by \cite{2014A&A...567A..75L}. Some errors are related to the telescope array, for example baseline instabilities from the AT relocations, flexure, pupil drifts and pointing uncertainties. The active field guiding (Section \ref{sec:FieldStabilization}) and pupil control (Section \ref{sec:PupilControl}) already minimize several of these errors. However, we have not yet implemented a more advanced post processing of, e.g., the acquisition camera data to correct for guiding residuals, and we have also not yet included the calibration of the narrow-angle astrometric baseline. Other error sources are related to the instrument, for example errors in the fiber positions, the dispersion of the single mode fibers and integrated optics, and the uncertainty in the wavelength calibration. We already include the dispersion in our analysis, we have made significant progress in the wavelength calibration, and we are currently exploring possibilities to improve the fiber positioning and object acquisition. The last family of errors mentioned here are related to the laser metrology, and the effects of non-common path aberrations and polarization on the path-length measurement. Also here we are still in process of transferring the theoretical concepts into the actual data analysis and calibration. 

The upcoming commissioning runs will concentrate on these aspects to further improve GRAVITY's narrow-angle astrometry towards the goal of $\rm 10\,\mu$as accuracy. 


\section{Summary}

GRAVITY and the VLTI set new standards in optical/infrared interferometry. Our first observations demonstrate fringe-tracking on unresolved stars as faint as $m_K \approx 10$\,mag, coherent exposures on objects fainter than $m_K \approx 15$\,mag, limiting magnitudes of $m_K \approx 17$\,mag, visibility accuracies better than 0.25\,\% and closure phase accuracy better than $0.5^\circ$, multi-wavelength interferometric imaging with a spectral resolution of $R \approx 4500$, spectro-differential astrometry with a few\,$\mu$as precision, and dual-field astrometry with 50\,$\mu$as residuals. 

The breakthrough with GRAVITY -- especially with respect to the observations of faint objects -- is best illustrated with the comparison to previous limits: before GRAVITY, the so far faintest object observed with near-infrared interferometry -- as part of the technical Dual-Field Phase-Referencing instrument demonstration with the Keck telescopes -- was $m_K = 12.5$\,mag \citep{2014ApJ...783..104W}, a factor 10 brighter than the flare from the Galactic Center black hole discussed in the previous chapter. The difference is even a factor 100 when compared with previously published science observations, for which the faintest objects had an apparent brightness around $m_K \approx 10.1$\,mag \citep{2012A&A...541L...9W} and $m_K \approx 10.2$\,mag \citep{2011A&A...527A.121K} with the VLTI and Keck Interferometer, respectively. For fringe-tracking -- a prerequisite for long-exposures and high spectral resolution interferometry -- GRAVITY is increasing the sensitivity of the VLTI by a factor 10 compared to what is possible with FINITO, for which the H-band limiting magnitude for phase-tracking is $m_H \approx 7.5$\,mag. The limiting magnitude records of GRAVITY make a whole new range of objects accessible to optical/infrared astronomy. 

At the same time GRAVITY can be used without specific interferometry expertise, is coming with a science grade data reduction software, and is offered to the world-wide community through the European Southern Observatory. Although the Galactic Center is the key science target for GRAVITY, the instrument allows the study of many other objects, both with the ATs and UTs. 

This paper demonstrates the unique capabilities and performance of GRAVITY with a number of notable firsts, including the first observation of the Galactic Center supermassive black hole and its fast orbiting star S2 with infrared interferometry, the first interferometric observations of a T\,Tauri star and a HMXB at high spectral resolution, the first 10\,$\mu$as spectro-differential interferometry of a quasar broad line region, and the first dual-field interferometric astrometry with residuals as low as 50\,$\mu$as.


\begin{acknowledgements}

Based on observations made with ESO Telescopes at the La Silla Paranal Observatory under programme IDs 60.A-9102 and 099.B-0162. We thank the technical, administrative and scientific staff of the participating institutes and the observatory for their extraordinary support during the development, installation and commissioning of GRAVITY. The instrument relies strongly on the VLTI infrastructure, and we want to especially thank for its timely upgrade to prepare for GRAVITY. Since July 2008, the work on the spectrometers of GRAVITY is supported in part by the German Federal Ministry for Education and Research (BMBF) under the grants Verbundforschung \#05A08PK1 \#05A11PK2 \& \#05A14PKA. G.~Perrin and K.~Perraut acknowledge the support of Agence Nationale de la Recherche contract \#ANR-06-BLAN-0421, LabEx OSUG@2020 (Investissements d’avenir  – ANR10LABX56), Action Sp\'ecifique ASHRA of CNRS/INSU and CNES, Action Spécifique GRAM of CNRS/INSU-INP and CNES. S.~Gillessen and C.~Deen acknowledge the support from ERC starting grant No. 306311. O.~Pfuhl acknowledges the support from ERC synergy grant No. 610058 “BlackHoleCam: Imaging the Event Horizon of Black Holes”. S.~Lacour acknowledges support from ERC starting grant No. 639248. J.~Sanchez-Bermudez acknowledges the support from the Alexander von Humboldt Foundation Fellowship programme (Grant number ESP 1188300 HFST-P). The Portuguese participation in GRAVITY was partially funded by Fundação para a Ciência e Tecnologia with grants PTDC/CTE-AST/116561/2010, COMPETE FCOMP-01-0124-FEDER-019965, UID/FIS/00099/2013 and SFRH/BD/52066/2012. This research was partially funded by European Community’s Seventh Framework Programme, under Grant Agreements 226604 and 312430 (OPTICON Fizeau Programme), and has made use of the Jean-Marie Mariotti Center \texttt{Aspro}, \texttt{OIFits Explorer} and \texttt{SearchCal} services and of CDS Astronomical Databases SIMBAD and VIZIER.

\end{acknowledgements}

\bibliographystyle{aa} 
\bibliography{GRAVITYFirstLight} 

\end{document}